\documentclass[12pt,preprint]{aastex}
\usepackage{CJK}

\usepackage{epsfig}
\usepackage{amsmath}
\usepackage{subfigure}
\usepackage{geometry}
\usepackage{graphicx}
\usepackage{float}
\usepackage{subfloat}
\usepackage{morefloats}

\newcommand{\kms}{\,km\,s$^{-1}$\,}

\newcommand{\msun}{\,$M_{\odot}$}
\newcommand{\msunyr}{\,\msun\,yr$^{-1}$\,}

\newcommand{\lsun}{\,$L_{\odot}$}

\shorttitle{ The spectra of evolved stars at 20--25\,GHz}
\shortauthors{Zhang et al.}

\begin{document}

\title{The spectra of evolved stars at 20--25\,GHz:
tracing circumstellar chemistry during the asymptotic giant branch to planetary nebula transition} 

\begin{CJK*}{UTF8}{gbsn}
\author{Yong Zhang$^{1}$, Wayne Chau$^{2}$,  Jun-ichi Nakashima$^{1}$, Sun Kwok$^{3}$}

\altaffiltext{1}{School of Physics and Astronomy, Sun Yat-Sen University, Tangjia, Zhuhai, China}
\altaffiltext{2}{Department of Physics,  University of Hong Kong, Pokfulam Road, Hong Kong, China}
%\altaffiltext{3}{Department of Astronomy and Geodesy, Ural Federal University}
\altaffiltext{3}{Department of Earth, Ocean, and Atmospheric Sciences, University of British Columbia, Vancouver, Canada}
%\altaffiltext{5}{Laboratory for Space Research, University of Hong Kong, Hong Kong, China}

\email{zhangyong5@mail.sysu.edu.cn}

\begin{abstract}
We report an unbiased radio line survey  towards the circumstellar envelopes of evolved stars at the frequency range from 20 to 25 GHz, aiming to obtain a more complete unbiased picture of the chemical evolution in the final stages of stellar evolution. The observation sample includes the asymptotic giant branch (AGB) star IRC+10216, the proto-planetary nebulae (PPNs) CRL\,2688 and CRL\,618, and  the young planetary nebula (PN) NGC\,7027, representing an evolutionary sequence spanning about 10,000 years. Rotational transitions from cyanopolyyne chains and inversion lines from ammonia are detected in the AGB star and PPNs, while the PN displays several  recombination lines. The different spectral behaviors of these evolved stars clearly reflect the evolution of circumstellar chemistry during the AGB-PPN-PN transitions.

%We present the observational results from a series of unbiased radio line surveys targeting the circumstellar envelopes of evolved stars. As part of an on-going effort, these surveys aim to expand the coverage both in terms of objects at different evolutionary stages and wavelength ranges in order to paint a more complete picture in our understanding of the chemical evolution for objects in the final stages of stellar evolution. The asymptotic giant branch star IRC+10216, proto-planetary nebulae CRL\,2688 and CRL\,618 as well as the planetary nebula NGC\,7027 were surveyed in the 20--25\,GHz frequency range using the 45-m telescope at the Nobeyama Radio Observatory. Rotational transitions from cyanopolyyne chains, ammonia absorption features and recombination lines are detected in the spectra of IRC+10216, CRL\,2688, CRL\,618 and NGC\,7027, respectively. In particular, the rotational transitions of HC$_7$N in CRL\,2688 and the recombination features on the spectra of NGC\,7027 have been clearly identified. Via comparing the results in the present paper with those presented in previous studies in our series of surveys, we were able to trace the chemistry and physical conditions within the circumstellar envelope as objects proceed on the evolutionary track.

\end{abstract}


\maketitle
\end{CJK*}

\section{Introduction}
\label{sec:intro}

The circumstellar envelope (CSE) of evolved stars comprises of materials that are ejected by the dying star. Observations at radio wavelengths have shown
that CSEs are the synthesis site of various gas molecules which are building
blocks for more complex organic compounds.
Due to the evolution of the central star and variation of mass loss, the physical conditions in the CSE --- such as density, temperature, and radiation field --- change rapidly as an object evolves from the asymptotic giant branch (AGB) to the proto-planetary nebula (PPN) and subsequently to the planetary nebula (PN) stage. Such environmental changes facilitate the onset of different chemical reactions that spawn a plethora of molecular species. The brevity of the PPN and PN phases, $\sim\!\!10^3$ and $\sim\!\!10^4$ years respectively, places excellent constraints on chemical models \citep{kwo04}. Understanding of the circumstellar chemistry must rely on the identification analysis of molecular features in objects at different evolutionary stages, making unbiased spectral line surveys the most ideal method to investigate conditions and composition of CSEs on a global scale \citep[see][for a comprehensive review]{cer11}. However, such studies are very time-consuming, and thus have mostly focused on a few brightest objects or been performed over limited frequency ranges. 

Recent technological advances in radio receivers have motivated increasing 
interests in performing unbiased line surveys towards CSEs, which
deliver an unbiased view of the circumstellar chemistry. The observations
using the same instrumental settings  minimize systematic uncertainties
and enable us to make a direct comparison between CSEs with different 
properties and at different evolutionary stages.  For instance,
\citet{ten1,ten2} carried out a 1\,mm spectral line survey of the 
O-rich CSE VY Camis Majoris and the C-rich CSE IRC+10216. These observations
revealed unexpected chemical complexity in O-rich environments \citep{ziu07}.
More recently, \citet{dan12}, \citet{gon15} and \citet{zha17} have conducted molecular line studies on IRC+10216 using the Herschel Space Observatory, the Effelsberg 100-m radio antenna and the Tian Ma Radio Telescope, respectively. In particular, the survey of \citet{gon15}, which covered the 17.8 -- 26.3\,GHz frequency range, detected a total of 78 transitions from known molecular species, amongst which 23 were detected for the first time outside the solar system. Our research group has been engaged in a long-term project to perform systematic spectral line surveys in a large sample of evolved stars with comprehensive wavelength coverage \citep{he08,zha08,zha09a,zha09b,cha12,zha13}. In the present paper, we investigate the 20--25\,GHz spectra
of a sample consisting of C-rich CSEs including IRC+10216, CRL\,2688, CRL\,618, and NGC\,7027. Displayed in 
Table~1 are the physical properties of the scientific objects consolidated from a series of literature searches. Unless otherwise stipulated, all analysis in this study will be based on the parameters listed in the table.

With over 70 molecular species discovered, IRC+10216 is often considered to be 
an archetypal C-rich AGB \citep[see, e.g.,][]{cer00,cer10,he08,pat11}. 
Although its chemical uniqueness is certainly a matter for debate, the 
spectrum of IRC+10216 serves as a useful baseline to compare and contrast with the spectra of the other objects in studying circumstellar chemistry.
CRL\,2688 is a PPN that left the AGB phase approximately 200--350\,years ago \citep{jur90,uet06}. Its central star has an effective
temperature of about $7.25\times10^3$\,K, and is not hot enough
to photoionize the CSE. Near-infrared (NIR) imaging studies
show that the multipolar morphology of CRL\,2688 originates from high-velocity outflows of H$_2$ \citep{lat93,sah98}, and 
interferometric observations reveal a series of collimated CO jets that are highly correlated with the H$_2$ emission
\citep{cox00}. The geometry of CRL\,2688 might introduce some chemical 
complexity. The dust- and shock-induced chemistry probably take place
in the central dusty torus and the polar direction, respectively.
A wide variety of molecules --- such as the silicon-bearing molecules SiO and SiS, cyanopolyyne chains up to HC$_9$N and carbon radicals up to C$_6$H --- have been detected in  CRL\,2688 \citep{luc86,tru93,bac97a,bac97b}. Line
surveys of this objects have been recently performed at 1\,mm, 1.3\,mm, and 3\,mm windows \citep{par08,zha13}.
CRL\,618 is a more evolved PPN having a central star with a higher
effective temperature \citep[$3.0\times10^4$\,K;][]{san04}. It is believed
to be on the verge of becoming a full-blown PN.  
Its morphology can be characterized by two components: a slowly
 expanding (19\kms) halo component that extends to over 15--20{\arcsec} with a mass of a couple of solar masses \citep{phi92,fuk94,san04} and a high-velocity ($\pm$300\,\kms) bipolar component about 3--5{\arcsec} in length that is significantly less massive \citep{ner92,tra02,bal13}. Previous line surveys of this object
cover a frequency range from 80--276\,GHz \citep{buj88,par04,par05,par07}.
NGC\,7027 is a young PN with a hot central star \citep[$\sim$2.19$\times10^5$\,K;][]{zha05}. 
Surprisingly, observations show that most of the molecules can survive 
photodissociation by the strong UV radiation field \citep{ed13}.
The similarity of species between diffuse clouds and PNs suggests 
that the ``survival'' molecules are dispersed into the interstellar medium 
(ISM) with the expansion of the nebulae \citep{zi06}. A comparison between
the spectral properties of the four typical targets in different
evolutionary stages of evolved stars can provide a more complete picture
of C-rich circumstellar chemistry.

%In this paper, we extend our series of surveys to cover the 20--25\,GHz frequency region for CRL\,2688, CRL\,618 and NGC\,7027. This facilitates the detection of molecular features belong to cyanopolyyne molecules and ammonia in the two PPNe as well as the identification of hydrogen and helium recombination lines in NGC\,7027. Although many of these lines have been observed by other authors, the present study offers the opportunity to compare and contrast the different spectra obtained using the same facility and under the same instrumental settings such that systematic uncertainties are minimized. The remainder of this paper is organized as follows. Section\,\ref{sec:obs} provides a brief account of the observational details. Observational results regarding the line features identified are reported in Section\,\ref{sec:results}. Implications of our observational results on circumstellar chemistry are discussed in Section\,\ref{sec:anal}. Finally, Section\,\ref{sec:summ} provides a summary and lays out the concluding remarks.

\section{Observation and Data Reduction}
\label{sec:obs}

The observations were carried out with the 45-m radio telescope at the Nobeyama Radio Observatory (NRO) during two period: May 2010 (for IRC+10216)
and February to May 2013 (for CRL\,2688, CRL\,618, and NGC\,7027).
Data were taken using the position switching mode, with on-source integration time of about 1--2 hr for each target. The receiver backend  is the 
digital spectrometer SAM45, which provides 16-array outputs, each with 4096 
spectral channels. The bandwidth and the channel spacing are 2\,GHz and
448\,kHz respectively. Two adjacent spectral channels were binned together,
to enhance the signal-to-noise ratio, resulting in a resolution of 
$\sim1$\,MHz or 12\,km\,s$^{-1}$ at 25\,GHz.  The beam size, $\theta_b$, and the main-beam efficiency, $\eta_m$, referenced at 23\,GHz are 
 73{\arcsec} and 0.83, respectively.
Typical system temperatures during the observations were between 90 and 180\,K with instances of it increasing to over 200\,K when weather conditions became less optimal. The pointing accuracy was checked every 2--3 hr
by observing nearby SiO maser sources. 
Spectra obtained were reduced using the NewStar\footnote{http://www.nro.nao.ac.jp/nro45mrt/obs/newstar} software package. The baselines were subtracted using
low-order polynomial fits to the line-free regions of the spectra. NGC\,7027,
a source with strong continuum emission, shows very irregular baseline curves in a few spectral 
regions. These `bad' spectral regions were discarded for further analysis.
Co-adding of the spectra resulted in a typical rms level of $\sim$5\,mK 
with a velocity resolution 12\,km\,s$^{-1}$
for IRC+10216 and CRL\,2688. The rms noises are somewhat higher for CRL\,618 and NGC\,7027 due to the shorter cumulative integration time. 

\section{Observational Results}
\label{sec:results}

\begin{figure}
	\centering
	\includegraphics[width=\textwidth]{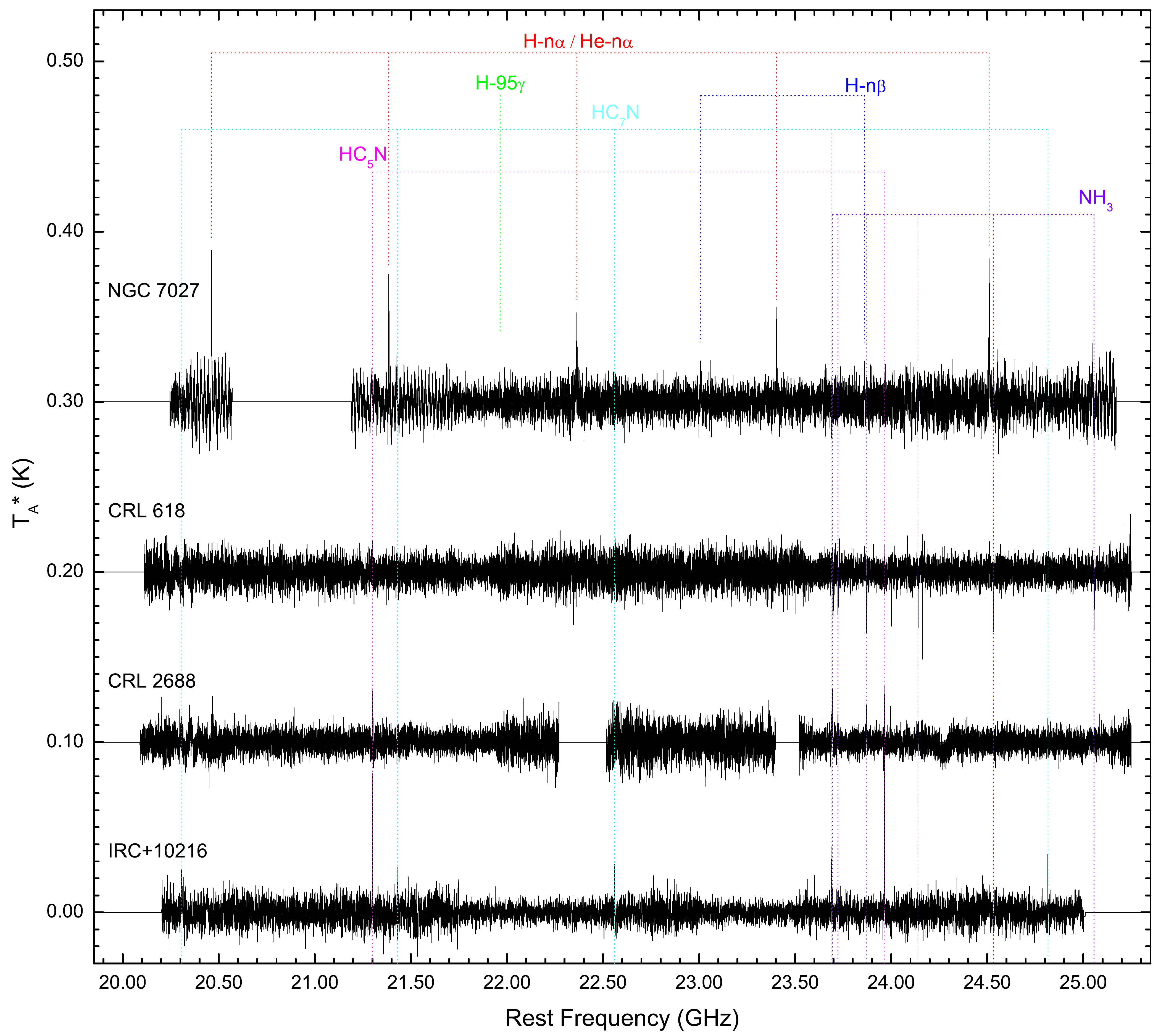}
	\caption{The 20--25\,GHz spectra of IRC+10216, CRL\,2688, CRL\,618 and NGC\,7027. Each subsequent spectra have been shifted upward by 0.1\,K. Spectra of CRL\,618 and NGC\,7027 had been scaled by a factor of 0.8 and 0.5 respectively to match the rms of the other spectra. Identified spectral features are indicated accordingly. Note that
the spectral regions with irregular baselines have been discarded.}	
	\label{fig:full}
\end{figure}

Figure~\ref{fig:full} displays the reduced spectra of
IRC+10216, CRL\,618, CRL\,2688, and NGC\,7027, while the zoomed-in spectra of the four objects 
are presented in Figure~\ref{fig:zoom}. These spectra are arranged in an evolutionary 
sequence from AGB to PPN to PN phases so that one can directly compare and contrast between 
spectra of objects at different dynamical ages. A total of 37 spectral features, including molecular emission and absorption as well as recombination lines, are positively detected in the four
objects.
Line identification relied on cross-referencing the spectra with various catalogues of transition data. Catalogues referenced by the present study include the National Institute of Standards and Technology (NIST) Recommended Rest Frequencies for Observed Interstellar Molecular Microwave Transitions\footnote{\url{https://physics.nist.gov/cgi-bin/micro/table5/start.pl}} \citep{lov04}, the JPL Catalogue of Molecular Spectroscopy\footnote{\url{https://spec.jpl.nasa.gov/ftp/pub/catalog/catform.html}} \citep{pic98} and the CDMS catalogue\footnote{\url{http://www.astro.uni-koeln.de/cgi-bin/cdmssearch}} \citep{mul01,mul05}. As a general rule of thumb, positive detection refers to features that have a peak intensity that 
is at least three times the rms level of its immediate spectral vicinity.  Features having the signal-to-noise of 1--2 are regarded as marginal detections.
The presence of these marginal detections can be argued based on other merits such as the positive detection of other transitions 
from the same molecular carrier within the CSE as well as detection by previous studies.

%\subsection{C-rich AGB IRC+10216}
\label{sec:10216}

\begin{figure}
	\centering
	\begin{subfigure}
		\centering
		\includegraphics[width=0.49\textwidth]{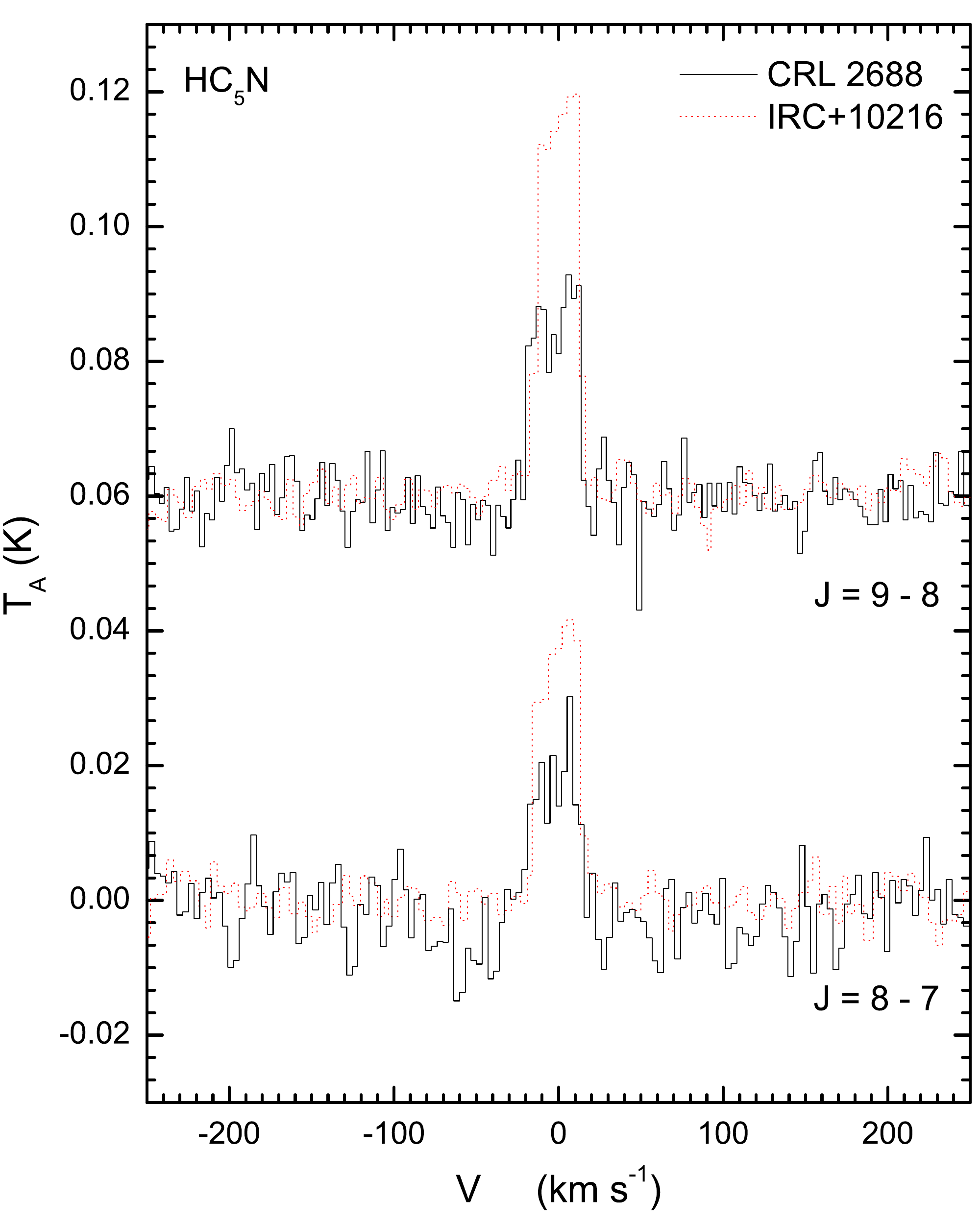}
	\end{subfigure}
	\begin{subfigure}
		\centering
		\includegraphics[width=0.49\textwidth]{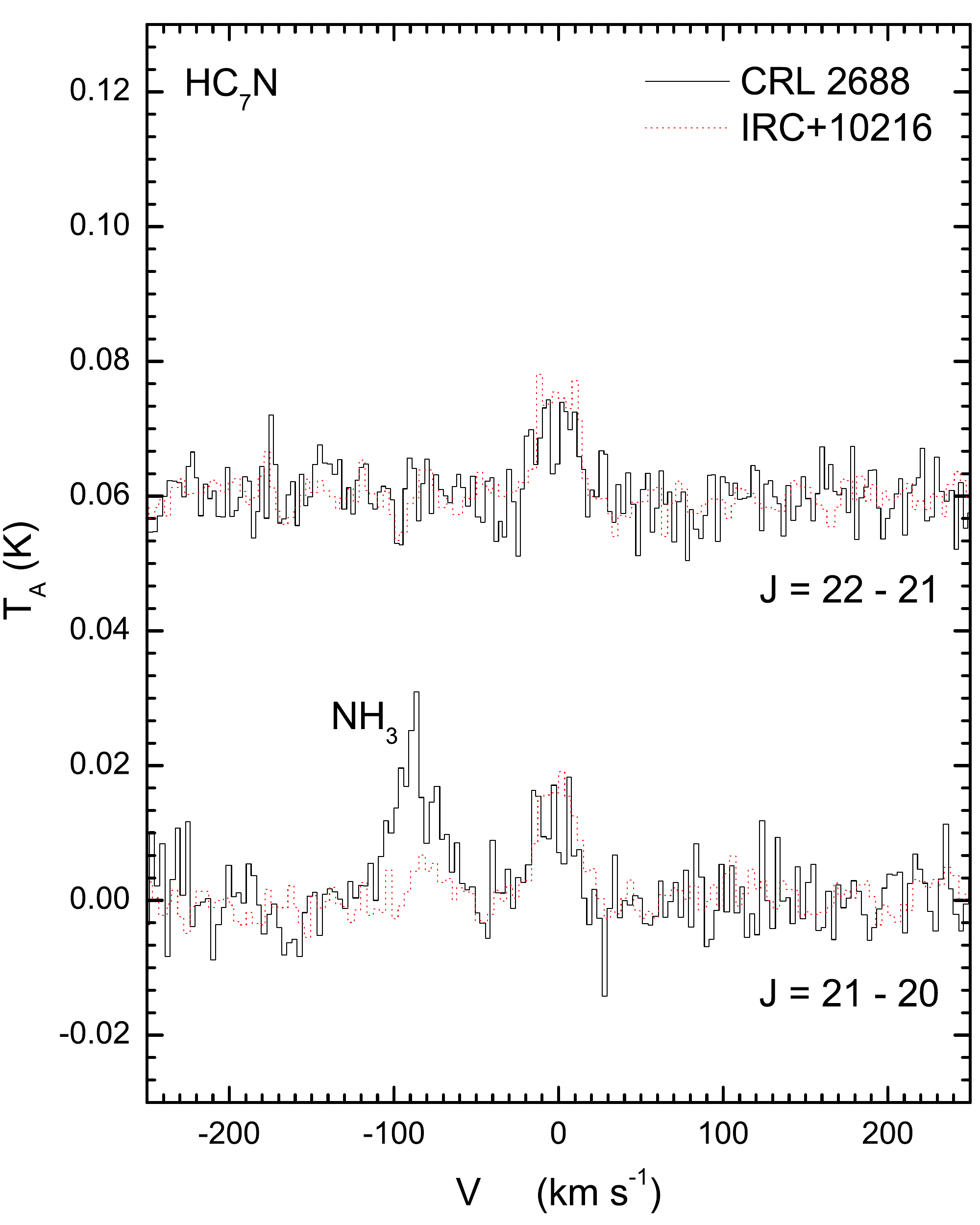}
	\end{subfigure}
	\begin{subfigure}
		\centering
		\includegraphics[width=0.49\textwidth]{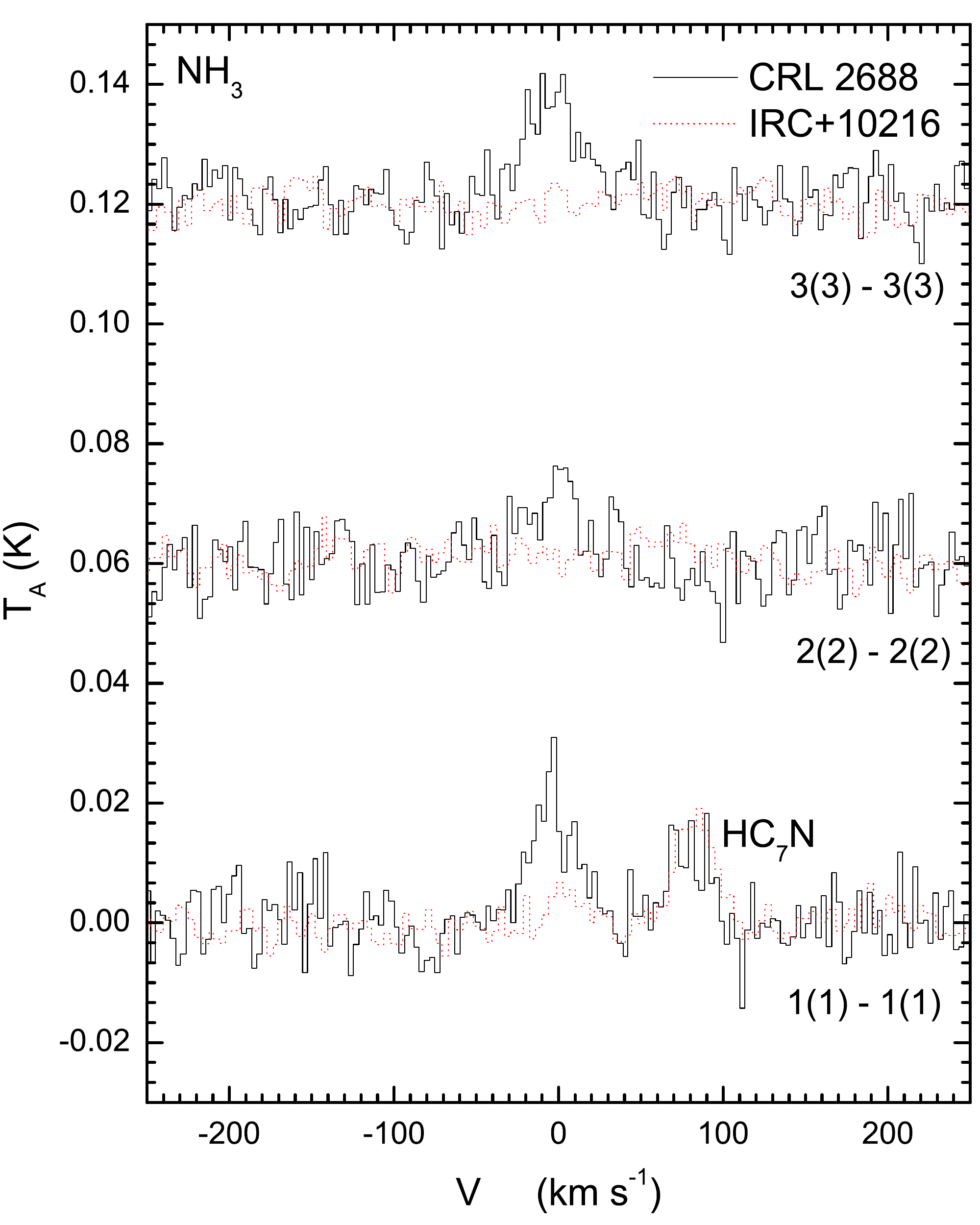}
	\end{subfigure}
	\caption{Line profiles of the prominent lines detected in the 20--25\,GHz spectrum CRL\,2688. The same lines identified in the spectrum of IRC+10216 have been scaled by 0.5 and superimposed on the CRL\,2688 profiles. 
{\bf The X-axis is given in $V_{\rm LSR}-V_{\rm SYS}$}.}\label{fig:2688_profiles}
\end{figure}

The NRO observation of IRC+10216 resulted in the detection of rotational lines from cyanopolyyne and ammonia spectral features, including the $J=8-7$ and $9-8$ lines from HC$_5$N, the $J=18-17$, $19-18$, $20-19$, $21-20$ and $J=22-21$ lines from HC$_7$N, as well as the $J-K=1-1$ from NH$_3$. The recent work can be 
regarded as an extension to \citet{kaw95} who performed a NRO observation of
IRC+10216 in the frequency range from 28--50\,GHz.
Measurements including peak and integrated intensities, line widths measured in full width at zero intensity (FWZI), and characteristic rms level are presented in Table~\ref{tbl:2688_lines}. 
The FWZI was measured from the distance in velocity space between the zero points of a given feature. 
Because of its sensitivity to line blending, fine-structure lines as well as the selection of zero points, the FWZI is not
necessary equal to two times of the terminal velocity.
From the table and the zoomed-in profiles in Figure~\ref{fig:2688_profiles}, it is apparent that the strength of the NH$_3$ $J-K=1-1$ feature is weak relative to those of the cyanopolyyne features. We did not detect  higher order ammonia transitions in IRC+10216. 
Generally, HC$_5$N  lines are stronger than HC$_7$N lines, and show less
symmetric profiles. As noted in \citet{cha12}, a similar trend, namely that the profile asymmetry is less for the larger molecule, is observed when comparing the HC$_3$N and HC$_5$N transitions at a higher frequency regime.

In previous studies, the longer cyanopolyyne chain HC$_9$N has been observed 
\citep[see, e.g.,][]{bel92, tru93}, four of which lie within our frequency 
range. However, HC$_9$N in IRC+10216 is less abundant than  HC$_7$N by one order of magnitude \citep{tru93}, and thus is not expected to be detected at  
our sensitivity level. Based on the observations performed with the
100-m radio telescope at Effelsberg, which has a HPBW of 36\farcs5 at 15\,GHz,
\citet{tru93} obtained that the HC$_9$N lines have a brightness temperature
of 11--16\,mK. Considering the beam-dilution effect, 
these lines in our spectra should have antenna temperatures of 
$T_{\rm A}<9$\,mK, well below the 3$\sigma$ detection limits.

%\subsection{PPN CRL\,2688}
\label{sec:2688}

The 20-25\,GHz spectrum of CRL\,2688 is quite similar with that of IRC+10216.
Emission lines from  HC$_5$N,  HC$_7$N, and NH$_3$are clearly detected.
However, their intensities are different from those in IRC+10216
(see Table~\ref{tbl:2688_lines} for a comparison). Comparing to IRC+10216,
CRL\,2688 exhibits stronger NH$_3$ lines, weaker cyanopolyyne lines, and
larger HC$_7$N/HC$_5$N intensity ratios.  Profiles of the most prominent lines in CRL\,2688 are presented in Figure~\ref{fig:2688_profiles} with their counterparts in the spectrum of IRC+10216 superimposed. The profiles associated with IRC+10216 have been scaled by a factor of 0.5. It is apparent that the HC$_5$N lines in CRL\,2688 show  a double-peak 
structure on which the red-ward peak is stronger, suggesting that these low-frequency 
HC$_5$N transitions are optically thin and primarily originate from the outer region of a 
detached envelope.   Such a behaviour
is typical for expanding circumstellare envelopes \citep[e.g.][]{cha12}.
The central dip is not seen in the HC$_5$N line profiles of
IRC+10216 in that this object has a smaller angular extent. 
The detected NH$_3$ features are groups of unresolved hyperfine structure of the $J-K=1-1,~2-2,~3-3$ and $4-4$ transitions. The detections of the $J-K=1-1$ and $2-2$ transitions in
CRL\,2688 have reported by \citet{ngu84}. In IRC+10216, though several 
NH$_3$ lines have been discovered previously \citep{gon15,sch16},
we only detect the strongest $J-K=1-1$ transition.
As shown in Figure~\ref{fig:2688_profiles}, 
the intensity of the NH$_3$ $J-K=1-1$ line is $\sim$63\% less than that of the HC$_7$N 
$J=21-20$ HC$_7$N line in IRC+10216, 
while its intensity is $\sim$72\% more than that of the same HC$_7$N line in CRL\,2688.
Together with the absence of other NH$_3$ transitions in IRC+10216, this ostensibly indicates that NH$_3$ is enhanced as an object evolves from the AGB to the PPN phase.

\begin{figure}
	\centering
	\includegraphics[width=0.49\textwidth]{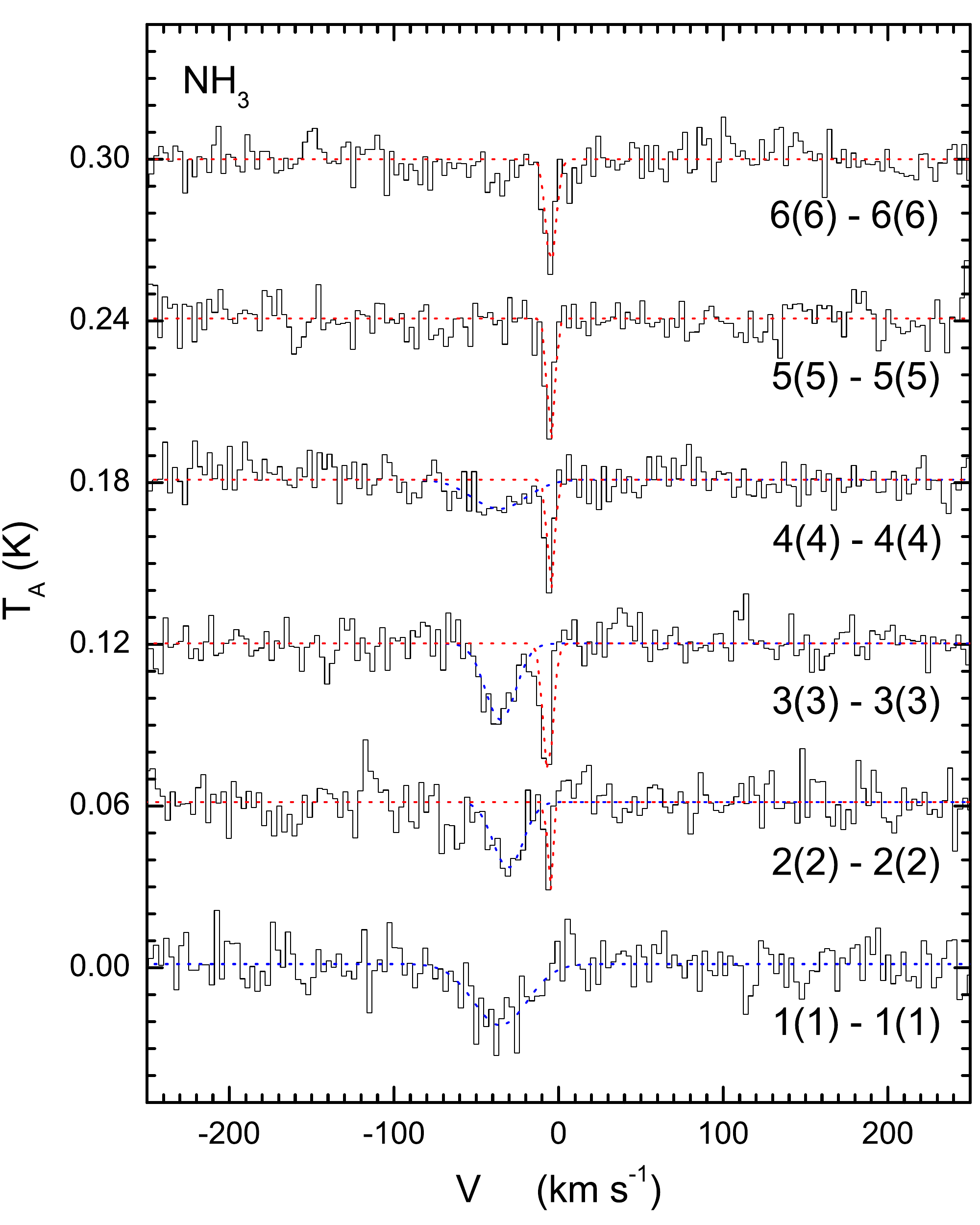}
	\caption{Profiles of NH$_3$ absorption features in the 20--25\,GHz spectrum of CRL\,618. Fittings of the broad (\textit{blue}) and narrow (\textit{red}) components are indicated by the dotted lines.
{\bf The X-axis is given in $V_{\rm LSR}-V_{\rm SYS}$.}
}\label{fig:618_profiles}
\end{figure}

\label{sec:618}
The 20--25\,GHz spectrum of  CRL\,618 reveals NH$_3$ absorption features of
transitions from $J-K=1-1$ to $6-6$.  No cyanopolyyne lines are detected with
our sensitivity. The profiles of these features are presented in 
Figure~\ref{fig:618_profiles}. All of these lines
are centered at negative velocities with respect to the local standard of rest,
and are composed of two components, wide and narrow ones. 
%suggesting that
%they originate from cold envelope completely detached from the central star.
The observations of NH$_3$ features in CRL\,2688 and CRL\,618 have 
been reported by \citet{tru88} and  \citet{mar92}. Our measurements are 
consistent with these prior observations.

%In the NRO spectrum of CRL\,618, only NH$_3$ absorption features have been detected, as apparent in Figure\,\ref{fig:full} and \ref{fig:zoom}. These absorption features are also composed of hyperfine structures that correspond to the same transitions as the NH$_3$ emission observed the spectrum of CRL\,2688 as well as those of the $J-K=5-5$ and $6-6$ transitions \citep{kuk70, kak75}. The profiles of these features are presented in Figure\,\ref{fig:618_profiles}. From profiles, it is observed that all of these lines are centered at negative velocities with respect to the local standard of rest. As these are absorption lines that arise when a cooler, optically thicker medium is between the observer and the source of continuum emission, they are indicative of an envelope that is completely detached from the central star. This shifting from NH$_3$ emission to NH$_3$ absorption between CRL\,2688 and CRL\,618, hence, supports the hypothesis that the latter is the more evolved amongst the two objects. 

%\subsection{PPN NGC\,7027}
\label{sec:7027}

\begin{figure}
	\centering
	\includegraphics[width=0.49\textwidth]{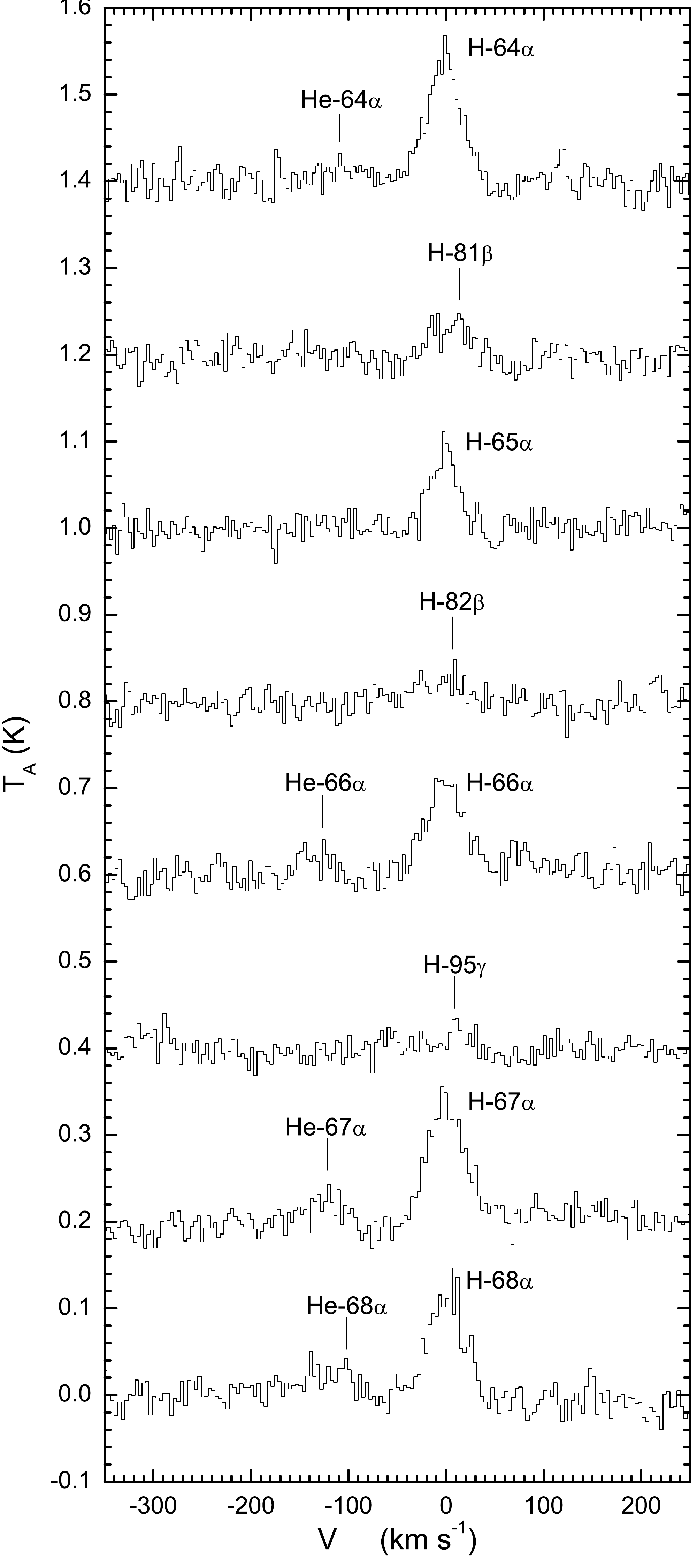}
	\caption{The profiles of hydrogen and helium recombination lines in the 20--25\,GHz spectrum of NGC\,7027. Profiles are arranged from top to bottom in terms of descending rest frequency. {\bf The X-axis is given in $V_{\rm LSR}-V_{\rm SYS}$}.}\label{fig:7027_profiles}
\end{figure}

The 20--25\,GHz spectrum of NGC\,7027 does not reveal any molecular species.
However, we detected strong recombination lines from hydrogen and much weaker recombination lines from helium.
Their rest frequencies, brightness temperatures, integrated intensities and line widths are listed in Table~\ref{tbl:recomb}. The line profiles are displayed in 
Figure~\ref{fig:7027_profiles}. The absence of molecular lines 
and the presence of recombination lines are presumably brought forth by the stronger 
radiation field resulting from the higher temperature of the central star.

%NGC\,7027 is the only PN surveyed in the present study. Although no molecular lines are detected in its 20--25\,GHz spectrum, a series of hydrogen and helium recombination lines are observed, as displayed in Figure\,\ref{fig:zoom}. The rest frequency, brightness temperature, integrated intensity and line width are listed in Table\,\ref{tbl:recomb} along with ther characteristic rms level of the region from which the recombination features are situated, while the line profiles are displayed in Figure\,\ref{fig:7027_profiles}. To the best of our knowledge, these lines have yet to be observed by previous studies on this PN. While the absence of molecular lines implies the photodissociation of molecules, the presence of recombination lines is indicative of an ionized region; both are brought forth by the stronger radiation field resulting from the higher temperature of the central stars of PNe. %As the higher stellar temperature is attributed to the stellar core being exposed via the ejection of the CSE, the observational results here are qualitatively consistent with the standard picture of late-stage stellar evolution.

\section{Analysis and Discussions}
\label{sec:anal}

\subsection{Cyanopolyynes}
\label{sec:cyano}

\begin{figure}
	\centering
	\begin{subfigure}
		\centering
%		\caption{(a) Rotational diagram of HC$_5$N.}
		\includegraphics[width=\textwidth]{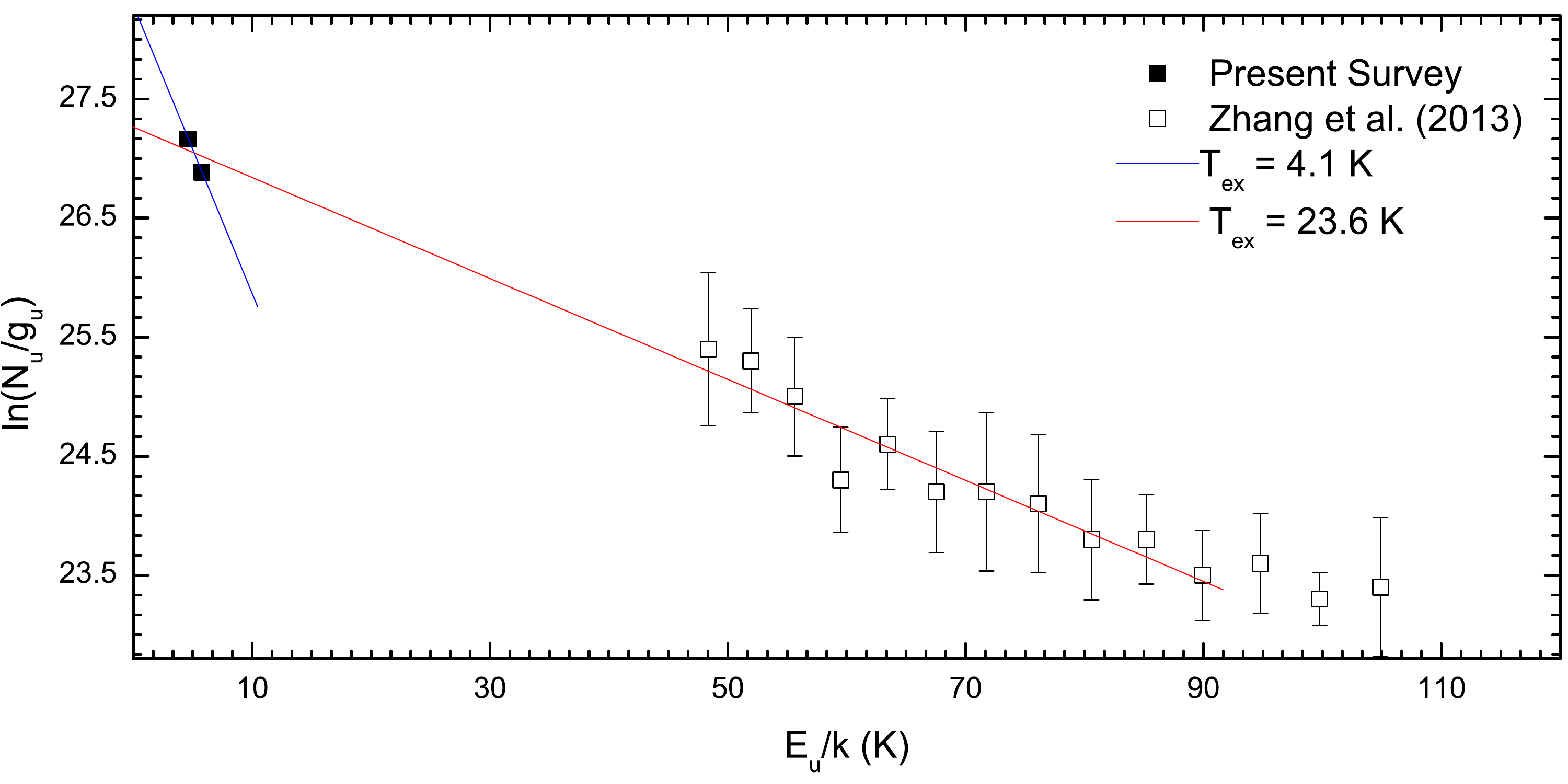}%\label{fig:2688_rot_hc5n}
	\end{subfigure}
	\begin{subfigure}
		\centering
%		\caption{(b) Rotational diagram of HC$_7$N.}
		\includegraphics[width=\textwidth]{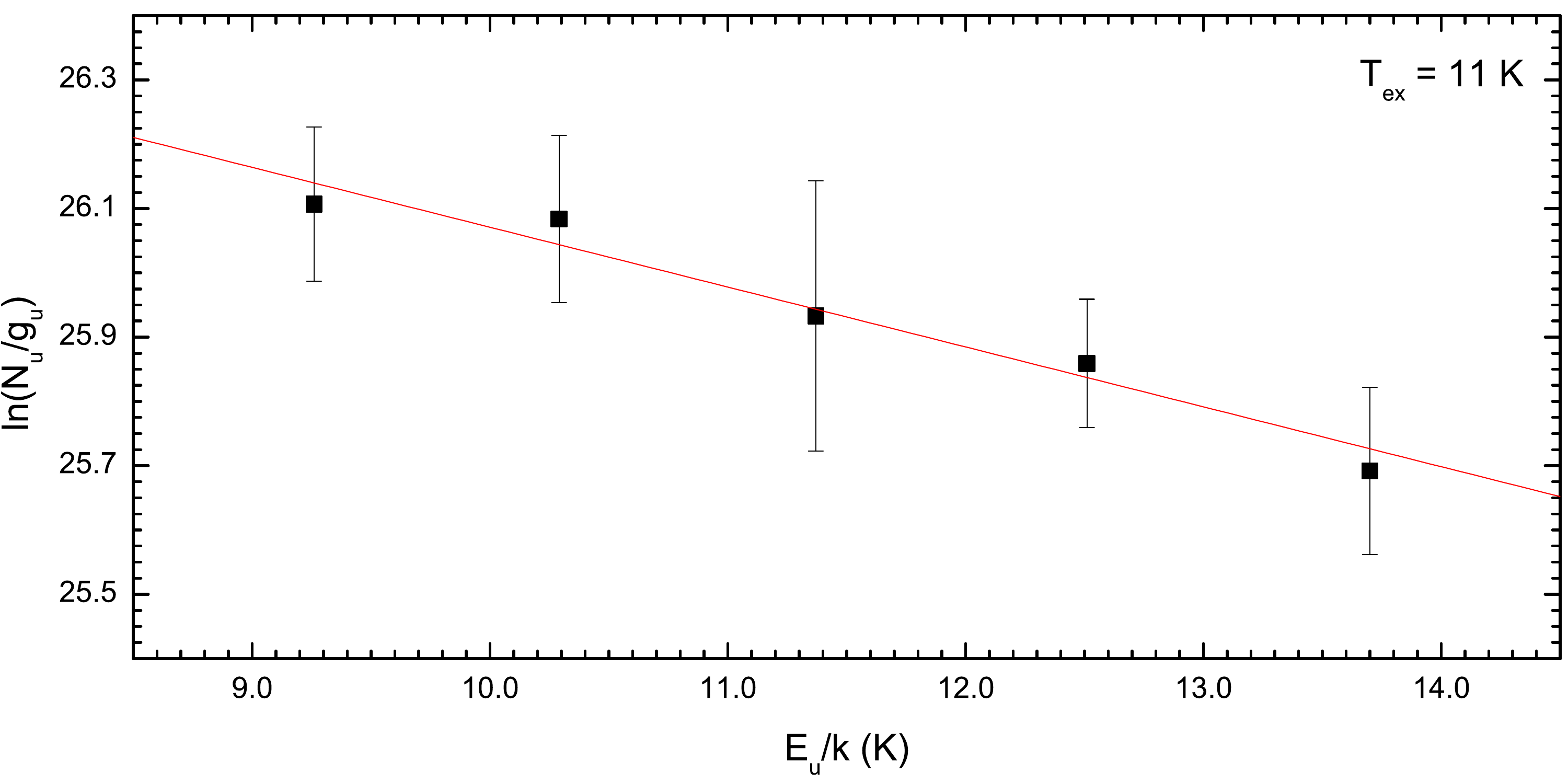}%\label{fig:2688_rot_hc7n}
	\end{subfigure}
	\caption{Rotational diagrams of cyanopolyyne molecules detected in the NRO spectrum of CRL\,2688. \textit{Upper panel:} Rotational diagram of HC$_5$N. Data points of HC$_5$N from \citet{zha13} have been concatenated with data from the present study. \textit{Lower panel:} Rotational diagram of HC$_7$N.}\label{fig:2688_rot}
\end{figure}

Cyanopolyyne is  a family of linear chain molecules with a CN radical end. The large dipole moment of 
the CN radical leads to intense rotation emission of cyanopolyyne molecules, and thus
the search for the cyanopolyyne species in interstellar and circumstellar environments has attracted great interests.  The longest cyanopolyyne molecule in space is HC$_{9}$N that was detected in IRC+10216.
%The longest cyanopolyyne molecule in space is HC$_{11}$N that was detected
%by \citep{bel82} in IRC+10216.
The relative abundances of cyanopolyyne compounds can provide important insights
on circumstellar chemistry \citep{par05}.
Cyanopolyyne molecules in the PPN CRL\,2688 are examined via rotational analysis. Under the assumption of optically thin and local thermal equilibrium (LTE), the population of the energy states are related to the integrated intensity through the Maxwell-Boltzmann distribution as follows:
\begin{equation}
	\ln{\frac{N_u}{g_u}}=\ln{\frac{3k\int{T_S\,\mbox{d}v}}{8\pi^3\nu\mu^2S}}=\ln{\frac{N}{Q}}-\frac{E_u}{kT_{\mbox{\scriptsize{ex}}}}\,,
	\label{eq:rot}
\end{equation}
\noindent where $\mu^2S$ is the square of the dipole moment multiplied by the line strength, $N_u$, $g_u$, and  $E_u$ are
the population, degeneracy, and excitation energy of the upper level,
$\nu$ is the line frequency, and $Q$ is the rotational partition function, which is in general a function of the excitation temperature, $T_{\mbox{\scriptsize{ex}}}$. Here, the integrated intensity $\int{T_S\,\mbox{d}v}$ refers to that of the source brightness temperature, $T_S$, which is computed from the antenna temperature by accounting for the main beam efficiency, $\eta$, and beam dilution using the following equation:
\begin{equation}
	T_S=\frac{1}{\eta}\,\frac{\theta_b^2+\theta_S^2}{\theta_S^2}T^*_{\rm A},
	\label{eq:dilution}
\end{equation}
\noindent where $\theta_b$ and $\theta_S$ are the half power beam width (HPBW) and angular size of the source, respectively. The $\eta$ value is taken to be 0.87 while $\theta_b$ is found to be 61\farcs3--66\farcs8 for the frequency range surveyed by linearly extrapolating values in \citet{kaw95}. With the integrated intensities obtained from the observational data, the left-hand-side of Equation\,(\ref{eq:rot}) can be plotted against $E_u/k$ to create rotational diagrams. Taking the inverse of the slope of the linear regression of such plot yields the excitation temperature.
Presented in Figure~\ref{fig:2688_rot} are the rotational diagrams of the cyanopolyyne molecules detected in CRL\,2688.  Since the present survey only covers 
the low energy range and the thermal structure within the CSE is often stratified, data points from previous high-frequency surveys \citep{zha13,cha12} have been concatenated with those from the present study for a more reliable estimate of the excitation temperatures and column densities. Results from rotational analysis of cyanopolyyne emission from CRL\,2688 and IRC+10216 are presented in Table~\ref{tbl:2688_rot}. {Also listed in the table are the fractional abundances with respect to H$_2$, which are computed using the formula given by \citet{olo96},
\begin{equation}\label{abundance}
f_{\rm X}=1.7\times10^{-28}\frac{v_e\theta_bD}{\dot{M}_{{\rm H}_2}}
\frac{Q(T_{ex})\nu_{ul}^2}{g_uA_{ul}}
\frac{e^{E_l/kT_{ex}}\int T_Rdv}{\int^{x_e}_{x_i}e^{-4\ln2x^2}dx},
\end{equation}
where  the expansion velocity $v_e$ is in km\,s$^{-1}$, the distance 
$D$ in pc and the mass loss rate  $\dot{M}_{{\rm H}_2}$ in M$_{\sun}\,{\rm yr}^{-1}$ 
can be obtained from Table~1, $\int T_Rdv$ is in K\,km\,s$^{-1}$,
the expansion velocity $v_e$ is in km\,s$^{-1}$, 
 $\nu_{ul}$  is the line frequency in GHz, $g_u$ is the statistical weight of
the upper level, $A_{ul}$ is the Einstein coefficient for the transition,
$E_l$ is the energy of the lower level, and $x_{i,e}=R_{i,e}/(\theta_bD)$
with $R_i$ and $R_e$ the inner radius and outer radius of the shell
taken from \citet{woo03}.  }

\begin{figure}
	\centering
	\includegraphics[width=\textwidth]{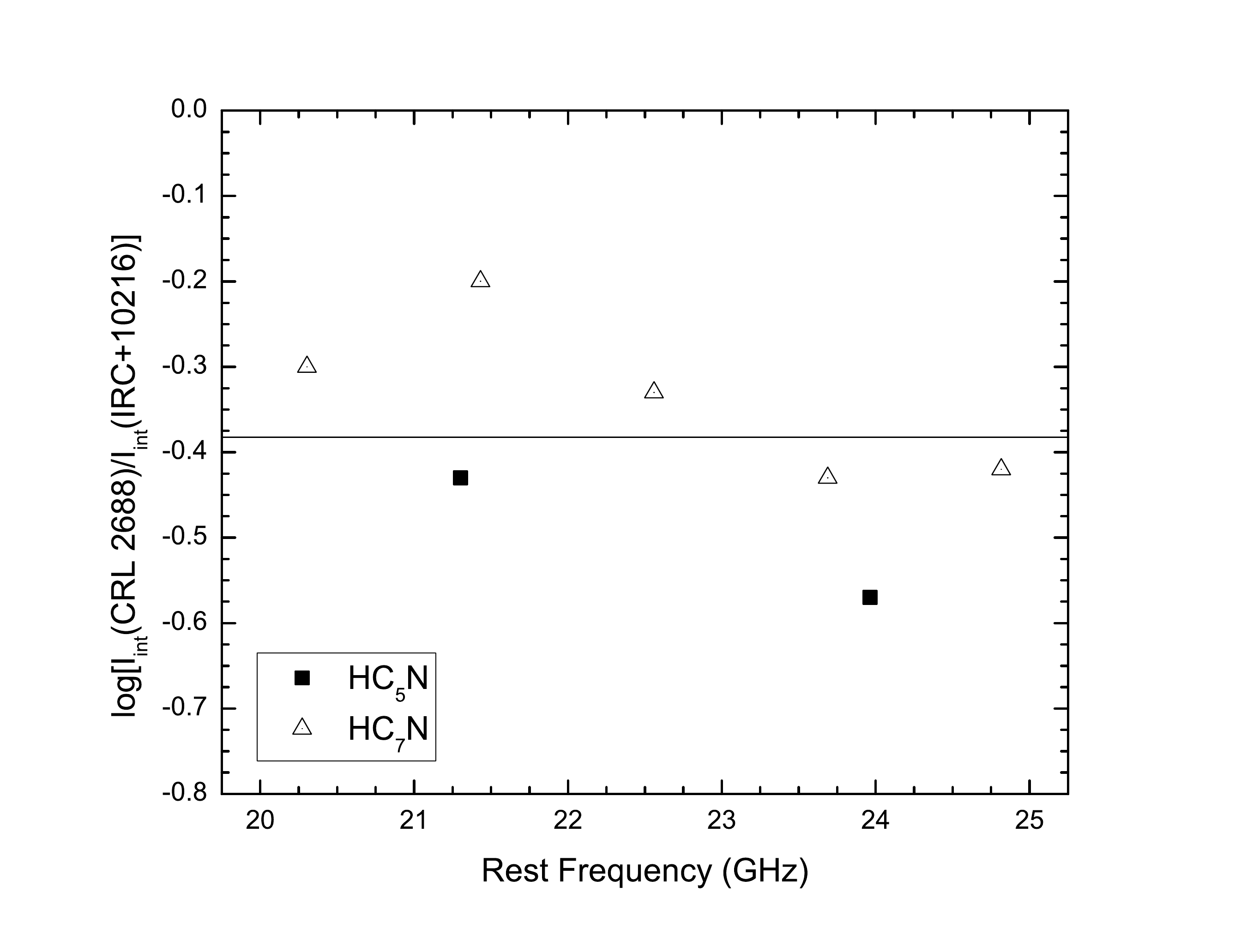}
	\caption{Integrated intensity ratios of detected cyanopolyyne molecules between CRL\,2688 and IRC+10216 after correcting for beam dilution. Ratios are presented on a logarithmic scale. The mean intensity ratio (0.41) is indicated by the solid line.}\label{fig:2688_ratios}
\end{figure}

In Figure~\ref{fig:2688_ratios}, we compare the 22--25\,GHz spectrum of CRL\,2688 with that of IRC+10216. Figure~\ref{fig:2688_ratios} plots the integrated intensity ratios between the two objects after adjusting for difference in the effects of beam dilution by a factor of $\left(1+\theta^2_b/\theta_{\mbox{\scriptsize{CRL\,2688}}}^2 \right)/\left(1+\theta^2_b/\theta_{\mbox{\scriptsize{IRC+10216}}}^2 \right)$ on a logarithmic scale. The angular size of the objects are taken to be $\theta_{\mbox{\scriptsize{CRL\,2688}}}=20$\arcsec~and $\theta_{\mbox{\scriptsize{IRC+10216}}}=30$\arcsec~\citep{fuk94}. As displayed in the figure, the integrated intensity ratios of the different transitions lie within the region from 0.27 to 0.63 and have a mean value of about 0.41,
in approximate agreement with the mean ratio of high-frequency lines
derived in \citet{zha13}.  There is evidence showing 
that the HC$_7$N/HC$_5$N ratio is enhanced in CRL\,2688 compared to 
IRC+10216, suggesting that the linear cyanopolyyne molecules have been quickly reprocessed to
 longer cyanopolyyne chains during the AGB-PPN transition.

% This figure is in good agreement with that derived in \citet{zha13} from a much larger collection of molecular species, implying that cyanopolyyne chains are neither enriched nor depleted in CRL\,2688. From a similar analysis for CIT\,6, it is observed cyanopolyynes are gradually enriched in the AGB stage \citep{cha12}. Such cyanopolyyne enhancement is clearly no longer present in the PPN stage. Although an overall cyanopolyyne enhancement is not observed in CRL\,2688, a relative enrichment in HC$_7$N, the larger of the two cyanopolyyne molecules detected, is apparent from Figure\,\ref{fig:2688_ratios}. This relative enhancement in the longer cyanopolyyne chain is consistent with the stepwise formation mechanism for these chain molecules.

%Since this is also the mean derived in \citet{zha13} for a much larger collection of molecular species, it is evident that cyanopolyyne chains are neither enriched nor depleted in CRL\,2688, and the cyanopolyyne enhancement in the AGB stage --- as seen in a similar analysis for CIT\,6 in \citet{cha12} --- is no longer present. This is consistent with the relative decrease in the fractional abundance of cyanopolyynes as an object evolves from the AGB to the PPN phase. Although an overall cyanopolyyne enhancement is not observed in CRL\,2688, a relative enrichment in HC$_7$N, the larger of the two cyanopolyyne molecules detected, is apparent from Figure\,\ref{fig:2688_ratios}. Such a relative enhancement in the longer cyanopolyyne chain is also consistent with the stepwise formation mechanism for these chain molecules.

\subsection{Ammonia}
\label{sec:ammo}

The detection of ammonia in circumstellar environments is intriguing as 
ammonia chemistry is possibly related to the formation of 
amino acids, the basic building blocks of life. 
Recent observations show that circumstellar ammonia is much richer than
theoretical predictions \citep[i.e.][]{wong18}.
In the present surveys,
ammonia is observed in IRC+10216 and CRL\,2688  in emission, as well as in 
CRL\,618 in absorption. As shown Figure~\ref{fig:2688_profiles},
the CRL\,2688 exhibits an enhancement of the NH$_3$/HC$_7$N ratio
compared to IRC+10216. Our observations are consistent with previous works
\citep{kwok81,ngu84,tru88,mar92}.  Although the existence of 
NH$_3$ in CSEs has been well established, its origin is not yet 
completely understood \citep{has06}. Interferometric observations of CRL\,2688
show that NH$_3$ emission is mostly confined in a central disk-like
structure, while HC$_7$N emission arises in an extended halo, suggesting
that the chemistry processes of the two molecules are different
\citep{ngu86}. In cold low-density environments, there is
no efficient gas-phase reaction to initiate the chemistry of N-bearing 
species. In current models of CSE chemistry \citet[][]{li16}, NH$_3$ is qualitatively 
assumed to have been formed in stellar atmospheres, and would be
processed into other N-bearing species in extended CSE.  The enhancement of NH$_3$ in CRL\,2688 suggests
that other mechanisms, such as shock-induced endothermic reactions and
solid surface  reactions, might be responsible for the formation of NH$_3$
during the AGB-PPN evolution.
It is indeed found that NH$_3$ emission is associated with shocked gas 
\citep{din09}. The initialization of nitrogen species chemistry 
in post-AGB stage can also be inferred by the high abundance of N$_2$H$^+$ in 
NGC\,7027 \citep{zha08}.  
Through a non-LTE modeling for NH$_3$ in IRC+10216, \citet{gon15} derived a 
kinetic temperature of 70\,K.
The NH$_3$(2,2)/(1,1) integrated ratio in CRL\,2688 is 0.68, a value 
very close to that in IRC+10216 obtained by \citet{gon15} (0.73). This
suggests that the NH$_3$ emitting regions in the two objects have 
a similar kinetic temperature.

%In the present series of surveys, ammonia is observed in three of the younger objects --- namely IRC+10216, CRL\,2688 and CRL\,618. The NH$_3$ lines in both IRC+10216 and CRL\,2688 are emission that originate from rotational transition. Such lines are much stronger in CRL\,2688, suggesting a relative enhancement of the molecule in the CSE towards the later stages of the late-stage stellar evolution. The fact that these NH$_3$ lines are absorption features indicates that such enrichment continues into the post-AGB phase.

%The presence of absorption lines implies that there exists, albeit much more compact than those in full-blown PNe, an ionized region providing the background continuum via free-free interaction. Since the flux ratio between free-free and bound-bound emission is weakly dependent on the electron temperature and number density, the strength of recombination features in CRL\,618 can be estimated from the intensities of NGC\,7027. From the \textit{Planck} spectral energy distributions in \citet{arn14}, it can be seen that the free-free flux ratio between CRL\,618 and NGC\,7027 should be about $9.22\times10^{-3}$. Accordingly, the peak intensities of hydrogen and helium recombination lines corresponding to those observed in NGC\,7027 should range from 0.28 to 1.57\,mK in CRL\,618. As this range is well below the rms level achieved in the present study, they are not detected in the NRO spectrum.

%The NRO spectrum of CRL\,618 in the present study is qualitatively consistent with that in \citet{mar92}. 

The NH$_3$ feature in CRL\,618 appears in absorption because there is
a compact ionized region in its center providing strong continuum emission
by free-free transitions. Following \citet{mar92}, we examined
the ammonia absorption features by assuming that they originate from two distinct components of the CSE. While the broad absorption features at approximately $-$55\kms are associated with a high-velocity wind, the much narrower components at around $-$27\kms are associated with a hot clump that is slower-moving. The narrow component is detectable in all NH$_3$ features except the one associated with the $J-K=1-1$ transition, and becomes more prominent for higher order transitions. By contrast, the broad component becomes less perceptible for increasing $J-K$ values to the point where it is barely visible in the $J-K=5-5$ and $6-6$ transitions as shown in Figure~\ref{fig:618_profiles}. From such observations, it is ostensible that the narrow component is from a much more energetic region that is capable of populating higher energy states that are left unpopulated in the region associated with the broad absorption component.
The broad and narrow NH$_3$ features in CRL\,618 are measured by fitting a Gaussian profile to each of the components, and the results  are presented in Table~\ref{tbl:618_abs}. 
Since the absorption features are yet to be saturated, the lines remain in the optically thin regime --- implying that the column density of each state is linearly proportional to the equivalent width. The column density of each $J-K$ level, $N_{J,K}$, can then be approximated from the equivalent width, $W_\nu$, using the equation \citep{draine}
\begin{equation}
       W_\nu \approx \left(\frac{1}{4\pi\varepsilon_0}\frac{e^2}{m_ec}\right)f_{lu}\times N_{J,K}\,,
       \label{eq:equiv_width}
\end{equation}
where $e$ is the electronic charge, $m_e$ is the mass of the electron, and $f_{ul}$ is the oscillator strength. 
The resulting column densities presented in Table~\ref{tbl:618_rot} are in
good agreement with  results of \citet{mar92}.

\begin{figure}
	\centering
	\includegraphics[width=\textwidth]{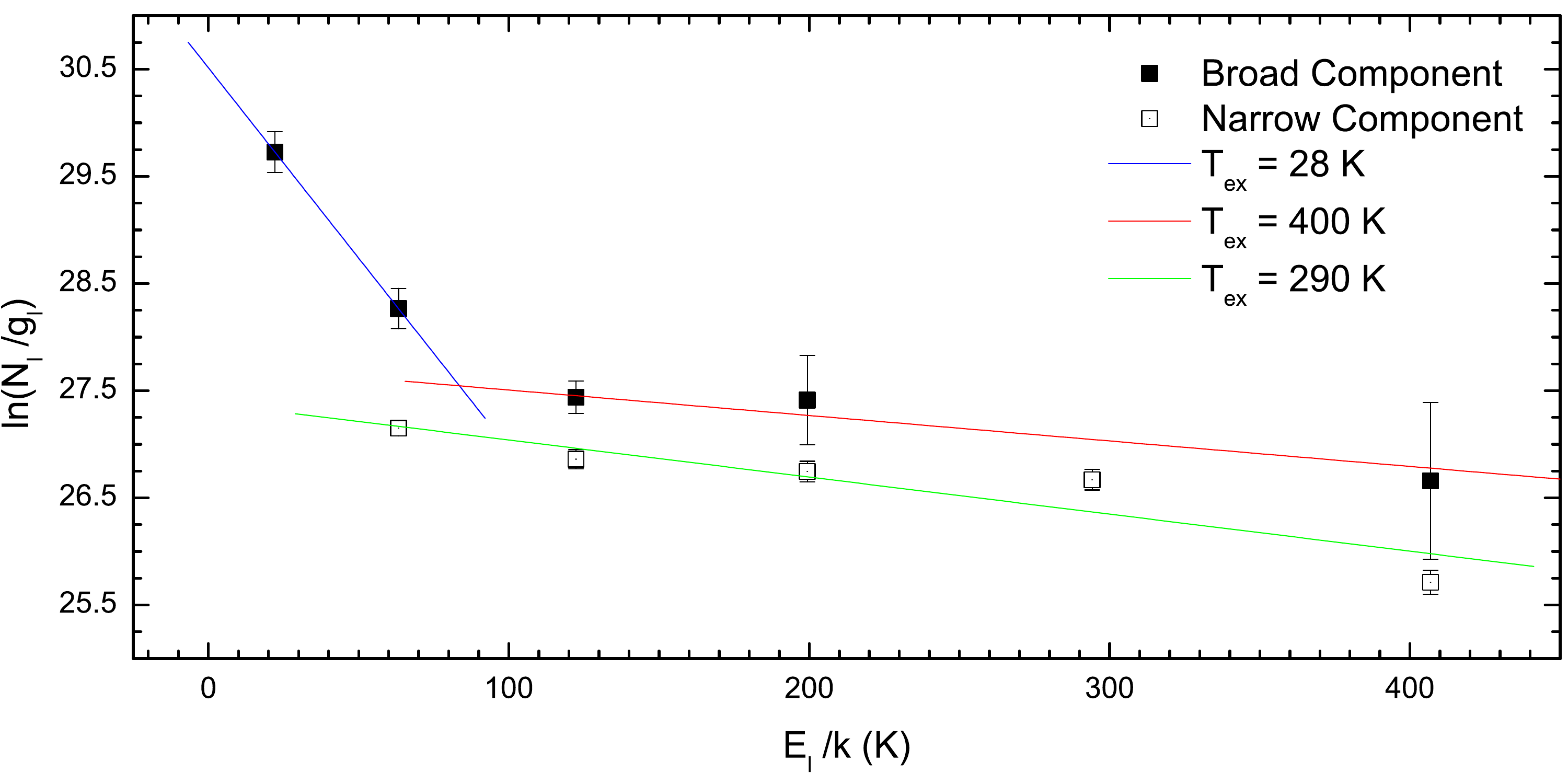}
	\caption{Rotational diagram for ammonia in CRL\,618. The broad absorption and the hot clump regions are treated as separate components. The broad absorption component cannot be characterized by a single excitation temperature.}\label{fig:618_rot}
\end{figure}

We also employed rotational diagram analysis to estimate the excitation temperature 
and the total column density of the NH$_3$ molecule in CRL\,618. 
Figure~\ref{fig:618_rot} presents the rotation diagram. 
It is evident from the figure that the broad absorption component cannot be characterized with a single rotation temperature. This suggests a stratification 
of the absorption region where the higher energy states are only populated in the warmer areas within the region. The narrow absorption feature is, on the other hand, well described by a single rotational temperature. Such characteristic supports the notion that the absorption region associated with the narrow features is compact in nature and can be well approximated by the LTE assumption.

Ammonia has been detected in diffuse clouds in absorption against bright
background sources \citep{lis06}.  Our sample observations suggest that
ammonia can be rapidly synthesized during the AGB-PPN transition. The
produced ammonia molecules then are gradually mixed with the diffuse ISM,
and can probably provide the starting materials for life, as suggested
by \citet{ziu09}.

\subsection{Recombination Lines}
\label{sec:recomb}

As presented in Figure~\ref{fig:7027_profiles}, a total of 12 hydrogen and helium recombination features are detected
or are marginally detected in the 20--25\,GHz spectrum of NGC\,7027. The peak and integrated intensities as well as the line width are listed in Table~\ref{tbl:recomb} along with the characteristic rms level in the region where the line resides. The mean widths of the H$n\alpha$, H$n\beta$ and He$n\alpha$ lines, namely $34\pm5$, $40\pm10$ and 43:\,\kms respectively, are found to be in reasonable
agreement with those associated with 
lower-$n$ transitions reported in \citet{zha08}.  The He$n\alpha$/H$n\alpha$ intensity ratio is
about $0.18\pm0.07$ , roughly consistent with the
He/H abundance ratio of this PN obtained by \citet{zha08}.

%He$^+$/H$^+$ abundance ratio has also been inferred by comparing the integrated intensities of the He$n\alpha$ and H$n\alpha$ lines of a given quantum number $n$. The resulting ratio of $N\left(\mbox{He}^+\right)/N\left(\mbox{H}^+\right)=0.19\pm0.07$ agrees well with that in \citet{zha08}.

%To further examine the recombination features, the integrated intensities are converted to measures of flux to facilitate cross comparison with theoretical models in \citet{sto95}. The main beam efficiency and flux conversion factor are taken to be 0.78 and 2.8\,Jy\,K$^{-1}$ throughout the frequency range of the present survey. Effects of beam dilution have already been accounted for via the flux conversion factor. Figure~\ref{fig:7027_ratio} plots the logarithm of the H$n\alpha$ to H39$\alpha$ ratio against the quantum number $n$. SMT 10-m and ARO 12-m telescope data of lower-$n$ transitions from \citet{zha08} have also been included in the figure to serve as a comparison. Uncertainties are estimated via differential error propagation form the observational data. Also presented in the figure are models from \citet{sto95} with various electron temperature and density assumptions under the Case~B approximation. It is evident from the figure that the observational data is consistent with the theoretical predictions for the electron temperature and density range specified and that the Case~B approximation, where downward transitions to the ground state may be ignored in the computation of statistic equilibrium, remains valid.

Recombination lines are not detected in CRL\,618 although this PPN
has a central ionized region. The spectral energy distributions
constructed by \citet{arn14} show that the  20--25\,GHz continuum emission 
of CRL\,618 is about 100 times fainted than that of NGC\,7027.
The flux ratio between free-free and bound-bound 
emission is only weakly dependent on the electron temperature and density.
Accordingly, if CRL\,618 and NGC\,7027 have the same line-to-continuum
ratio, the peak intensities of radio recombination lines in
CRL\,618 should be lower than 2\,mK, which is well below the rms level 
achieved in the present observations. Moreover, the central ionized region
of CRL\,618  is much more compact than that of NGC\,7027, and thus its
observations suffer from more severe effect from the beam dilution.
Therefore, it is extremely  difficult
to detect radio recombination lines in CRL\,618.

\section{Summary }
\label{sec:summ}

In this paper, we present an unbiased survey for IRC+10216, CRL\,2688, CRL\,618 and NGC\,7027 in the 20--25\,GHz frequency region, aiming to
obtain an unbiased view of circumstellar chemistry during the AGB-PPN-PN 
evolution.  Under our detection sensitivity, we do not detect new species and 
unexpected lines in the surveyed frequency range. Emission lines from 
cyanopolyynes and ammonia are detected in IRC+10216 and CRL\,2688.
The spectra of CRL\,618 and NGC\,7027 are dominated by  ammonia absorption features and bright radio recombination lines, respectively. From our observations, 
we can draw the following picture for circumstellar chemistry. During the 
AGB-PPN transition, the cyanopolyyne chains are continuously generated, while
ammonia is abruptly enhanced through a trigger that might be related to the
fast stellar winds of PPNs. At the onset of PPN-PN stage,
a considerable number of molecules carried by fast stellar winds
rapidly escape from the destruction of the increasingly harder UV 
radiation from the central stars.  These molecules gradually cool down, 
become invisible, and are dispersed into the ISM.

\acknowledgments
We are grateful to the anonymous referee for her/his constructive comments that contributed to improve the manuscript.
We thank Shuji Deguchi, Bosco~H.~K.~Yung, and the staff at NRO for their assistance during the observations. 
The 45-m radio telescope is operated by Nobeyama Radio Observatory, a branch of National Astronomical Observatory of Japan. 
This work was supported by National Science Foundation of China (NSFC, Grant No. 11973099), and a grant to SK from the Natural and Engineering Research Council of Canada.

\begin{deluxetable}{lcccc}
\tablecaption{Properties of the sources.}
\label{prop}
\tabletypesize{\scriptsize}
\tablewidth{0pt}
\tablehead{      & \colhead{IRC+10216}      & \colhead{CRL\,2688} & \colhead{CRL\,618} & \colhead{NGC\,7027}}
\startdata
%IRAS No.              & 09452+1330                             & \nodata                      & 04395+3601             & \nodata \\
%Alternate name                       & CW Leo                               & Egg Nebula                 & Westbrook Nebula     & \nodata \\    
Object type                            & AGB                           & PPN                             & PPN                            & PN \\
R.A.$^{\tiny\mbox{a}}$                                & 09:47:57.406                        & 21:02:18.27               & 04:42:53.64              & 21:07:01.593 \\
Dec.$^{\tiny\mbox{a}}$                             & +13:16:43.56                    & +36:41:37.0               & +36:06:53.4             & +42:14:10.18 \\
$P$\,(days)$^{\tiny\mbox{b,f}}$           & 630                            & 91                                & \nodata                     & \nodata \\
$L$\,(\lsun)$^{\tiny\mbox{b,g,k,l}}$                  & 9600                    & 3300                            & 10000                        &  8100 \\
$D$\,(pc)$^{\tiny\mbox{b,h,k,l}}$                     & 120                         & 420                              & 900                             &  880 \\
$T_*$\,(K)$^{\tiny\mbox{b,e,g,l,o}}$    & $2.5\times10^3$               &  $7.25\times10^3$         & $3.0\times10^4$       &  $2.19\times10^5$\\
$V_{\tiny{\mbox{LSR}}}$\,(\kms)$^{\tiny\mbox{c,e,j}}$    & $-$26.4             & $-$32                                & $-$24                             & 25 \\
$v_e$\,(\kms)$^{\tiny\mbox{c,d,i}}$              & 14.5             & 19         & 19                              &  30 \\
$\dot{M}$\,(\msunyr)$^{\tiny\mbox{c,d,i}}$           & $1.2\times10^{-5}$   & $1.4\times10^{-4}$    & $6.7\times10^{-5}$  &  $1.1\times10^{-4}$ \\
$\theta_s$(\arcsec)$^{\tiny\mbox{c,m,n}}$         &  30                           & 20                          & 20                         & 70
\enddata
\tablenotetext{a}{\,FK5 coordinate obtained from the SIMBAD Astronomical Database.}
\tablenotetext{b}{\,Pulsation period, stellar luminosity, distance, and stellar temperature of IRC+10216 obtained from \citet{woo03}.}
\tablenotetext{c}{\,LSR velocity, expansion velocity, mass-loss rate, and angular size of CRL\,2688, and CRL\,618 obtained from \citet{fuk94}.}
\tablenotetext{d}{\,Expansion velocity and mass-loss rate of IRC+10216 obtained from \citet{woo03}.}
\tablenotetext{e}{\,Stellar temperature and LSR velocity of IRC+10216 obtained from \citet{he08}.}
\tablenotetext{f}{\,Pulsation period of CRL\,2688 obtained from The International Variable Star Index.}
\tablenotetext{g}{\,Stellar luminosity and stellar temperature of NGC\,7027 obtained from \citet{zha05}.}
\tablenotetext{h}{\,Distance of NGC\,7027 adopted from \citet{mas89}.}
\tablenotetext{i}{\,Mass-loss rate, and expansion velocity of NGC\,7027 obtained from \citet{zha08}.}
\tablenotetext{j}{\,LSR velocity of NGC\,7027 adopted from \citet{hua10}.}
\tablenotetext{k}{\,Luminosity and distance of CRL\,2688 obtained from \citet{uet06}.}
%The distance previously adopted is closer to 1\,kpc; such change results from a new estimation of interstellar extinction in the direction of CRL\,2688 \citep[see][for details]{uet06}.}
\tablenotetext{l}{\,Luminosity, distance and temperature of CRL\,618 adopted from \citet{san04}.}
\tablenotetext{m}{\,Angular size of IRC+10216 and CRL\,618 adopted from \cite{fuk94} and \cite{san04}, respectively.}
\tablenotetext{n}{\,Angular size of the molecular emission region of NGC\,7027 obtained from \citet{bie91}. The ionized region of the PN is closer to 10\arcsec~\citep{has01}.}
\tablenotetext{o}{\,Stellar temperature of CRL\,2688 obtained from \citet{zha13}.}
\end{deluxetable}

\begin{deluxetable}{llcccll@{\extracolsep{0.1in}}cccc}
\rotate
\tablecaption{Molecular transitions detected in CRL\,2688 and IRC+10216 in the 20--25\,GHz frequency range.
\label{tbl:2688_lines}}
\tabletypesize{\scriptsize}
\tablewidth{0pt}
\tablehead{
\colhead{Frequency}& \colhead{Species} & \colhead{Transition} &  \multicolumn{4}{c}{CRL\,2688}  & \multicolumn{4}{c}{IRC+10216} \\
\cline{4-7}\cline{8-11}
 & & & \colhead{rms} &\colhead{$T^*_{\rm A}$} &\colhead{$\int T^*_{\rm A}\mathrm{d}v^\dagger$} & \colhead{${\Delta v_{\mathrm{FWZI}}}^\dagger$} & \colhead{rms} & \colhead{$T^*_{\rm A}$} &\colhead{$\int T^*_{\rm A}\mathrm{d}v^\dagger$} & \colhead{$\Delta v_{\mathrm{FWZI}}^\dagger$}\\
\colhead{(MHz)} &  &  & \colhead{(mK)} & \colhead{(K)} & \colhead{(K\,km/s)} & \colhead{(km/s)} & \colhead{(mK)} & \colhead{(K)} & \colhead{(K\,km/s)} & \colhead{(km/s)}}
\startdata
21301.3 & HC$_5$N & $8-7$ & 5 & 0.034 & 0.85$\pm$0.29 & 58.4 & 8 & 0.083 & 2.27$\pm$0.51 & 63.3 \\
23963.9 & HC$_5$N & $9-8$ & 6 & 0.033 & 0.88$\pm$0.22 & 36.5 & 8 & 0.119 & 3.23$\pm$0.55 & 68.8 \\
\\                     
20303.9 & HC$_7$N & $18-17$ & 6 & 0.014 & 0.32: & 50.5: & 6 & 0.025 & 0.64$\pm$0.26 & 42.8 \\
21431.0 & HC$_7$N & $19-18$ & 5 & 0.013 & 0.36: & 68.0: & 8 & 0.028 & 0.58$\pm$0.32 & 40.6 \\
22559.9 & HC$_7$N & $20-19$ & 11 & 0.016 & 0.36: & 48.8: & 5 & 0.028 & 0.77$\pm$0.42 & 83.7 \\
23687.9 & HC$_7$N & $21-20$ & 6 & 0.018 & 0.38$\pm$0.26 & 43.0 & 8 & 0.038 & 1.03$\pm$0.48 & 59.5 \\
24815.9 & HC$_7$N & $22-21$ & 6 & 0.014 & 0.37: & 53.2: & 5 & 0.036 & 0.96$\pm$0.34 & 68.9 \\
\\                     
23694.5 & NH$_3$ & $1\left(1\right)-1\left(1\right)$ & 6 & 0.031 & 0.74$\pm$0.46 & 77.2 & 8 & 0.014 & 0.26: & 55.7: \\
23722.6 & NH$_3$ & $2\left(2\right)-2\left(2\right)$ & 6 & 0.016 & 0.50:    & 55.6: & 8 & \nodata & \nodata & \nodata \\
23870.1 & NH$_3$ & $3\left(3\right)-3\left(3\right)$ & 6 & 0.022 & 0.75$\pm$0.40 & 67.4 & 8 & \nodata & \nodata & \nodata \\
24139.4 & NH$_3$ & $4\left(4\right)-4\left(4\right)$ & 6 & 0.012 & 0.43: & 78.9: & 6 & \nodata & \nodata & \nodata 
\enddata

\tablenotetext{\dagger}{\,Marginal detections are annotated with a colon (:).}

\end{deluxetable}

\begin{deluxetable}{cccclc}
%\rotate
\tablecaption{Hydrogen and helium recombination lines in NGC\,7027.
\label{tbl:recomb}}
\tabletypesize{\scriptsize}
\tablewidth{0pt}
\tablehead{
\colhead{Transition} & \colhead{Frequency} & \colhead{rms} &\colhead{$T_R$} & \colhead{$\int T_R\,\mbox{d}v^\dagger$} & \colhead{${\Delta v_{\mbox{\tiny FWHM}}}^\dagger$} \\
& \colhead{(MHz)} & \colhead{(mK)} & \colhead{(K)} & \colhead{(\kms K) } & \colhead{(\kms)} }
\startdata
H 64$\alpha$ & 24509.9 & 18 & 0.17 & 6.02 $\pm$ 1.72 & 33.9 \\
H 65$\alpha$ & 23404.3 & 13 & 0.11 & 2.98 $\pm$ 0.98 & 26.4 \\
H 66$\alpha$ & 22364.2 & 15 & 0.11 & 5.21 $\pm$ 1.67 & 39.8 \\
H 67$\alpha$ & 21384.8 & 21 & 0.16 & 6.45 $\pm$ 2.16 & 36.1 \\
H 68$\alpha$ & 20461.8 & 20 & 0.15 & 5.17 $\pm$ 1.86 & 35.7 \\
H 81$\beta$ & 23860.9 & 13 & 0.05 & 1.36 $\pm$ 0.76 & 34.5 \\
H 82$\beta$ & 23008.6 & 13 & 0.05 & 1.18 $\pm$ 1.03 & 52.1 \\
H 95$\gamma$ & 21964.3 & 13 & 0.03 & 0.60:   & 12.3: \\
He 64$\alpha$ & 24509.9 & 18 & 0.03 & 0.54: & 43.0: \\
He 65$\alpha$ & 23413.8 & 13 & $<0.02$ & $<0.80$   & \nodata \\
He 66$\alpha$ & 22373.3 & 15 & 0.04 & 0.85: & 28.4: \\
He 67$\alpha$ & 21393.6 & 21 & 0.04 & 1.02: & 23.1: \\
He 68$\alpha$ & 20470.1 & 20 & 0.05 & 1.52: & 77.2:
\enddata
\tablenotetext{\dagger}{\,Marginal detections are annotated with a colon (:).}

\end{deluxetable}

\begin{deluxetable}{lcccccc}
%\rotate
\tablecaption{Excitation temperatures, column densities and fractional abundances with respect to molecular hydrogen of identified cyanopolyyne molecules in CRL\,2688 and IRC+10216.
\label{tbl:2688_rot}}
\tabletypesize{\scriptsize}
\tablewidth{0pt}
\tablehead{
\colhead{Species} & \multicolumn{6}{c}{CRL\,2688}\\
\cline{2-7}
 & \colhead{$T_{ex}$} & \colhead{$\delta_{T_{ex}}$} &\colhead{$N$} & \colhead{$\delta_{N}$} &\colhead{$f_{X}$} & \colhead{$\delta_{f_{X}}$} \\
 & \colhead{(K)} & \colhead{($\pm$K)} & \colhead{($10^{14}$ cm$^{-2}$)} & \colhead{($\pm10^{14}$ cm$^{-2}$)} & \colhead{($\times 10^{-7}$)}  & \colhead{($\pm \times 10^{-7}$)}  }
\startdata
HC$_{5}$N$^a$ & 24  & 1 & 2.6 & 0.4 & 0.6 & 0.1 \\
HC$_{7}$N$^b$ & 11 & 1 & 0.24 & 0.03 & 0.4 & 0.01 \\
\cline{1-7}\\
\colhead{Species} & \multicolumn{6}{c}{IRC+10216}\\
\cline{2-7}
 & \colhead{$T_{ex}$} & \colhead{$\delta_{T_{ex}}$} &\colhead{$N$} & \colhead{$\delta_{N}$} &\colhead{$f_{X}$} & \colhead{$\delta_{f_{X}}$} \\
 & \colhead{(K)} & \colhead{($\pm$K)} & \colhead{($10^{14}$ cm$^{-2}$)} & \colhead{($\pm10^{14}$ cm$^{-2}$)} & \colhead{($\times 10^{-7}$)}  & \colhead{($\pm \times 10^{-7}$)}\\
\cline{1-7} \\
HC$_{5}$N$^c$ & 21 & 6 & 8 & 2 & 7  &  1 \\
HC$_{7}$N & \nodata & \nodata & \nodata & \nodata & 7 & 1
\enddata
\tablenotetext{a}{Concatenated with data from \citet{zha13}.}
\tablenotetext{b}{Computed with data points from the present survey alone.}
\tablenotetext{c}{Concatenated with data from \citet{cha12}.}
%\tablenotetext{d}{Inner and outer boundary of the emission region adopted from \citet{woo03b}.}
\end{deluxetable}

\begin{deluxetable}{cccccccccccc}
\rotate
\tablecaption{Broad and narrow ammonia absorption features detected in the 20--25\,GHz spectrum of CRL\,618.\label{tbl:618_abs}}
\tabletypesize{\scriptsize}
\tablewidth{0pt}
\tablehead{
\colhead{Transition} & \colhead{Frequency} & \multicolumn{10}{c}{Broad Absorption Features}\\
\cline{3-12}
 & & \colhead{$T_A^*$} & \colhead{$\delta_{T_A^*}$} & \colhead{$V_{\tiny{\mbox{LSR}}}$} & \colhead{$\delta_{V_{\tiny{\mbox{LSR}}}}$} & \colhead{$\Delta_{\tiny{\mbox{FWHM}}}$} & \colhead{$\delta_{\Delta_{\tiny{\mbox{FWHM}}}}$} & \colhead{Area} & \colhead{$\delta_{\tiny{\mbox{Area}}}$} & \colhead{$N_{J,K}$} & \colhead{$\delta_{N_{J,K}}$}\\
\colhead{$J-K$} & & \colhead{(mK)} & \colhead{($\pm$mK)} & \colhead{(km/s)} & \colhead{($\pm$km/s)} & \colhead{(km/s)} & \colhead{($\pm$km/s)} & \colhead{(K km/s)} & \colhead{($\pm$K km/s)} & \colhead{($10^{12}\mbox{cm}^{-2}$)} & \colhead{($\pm 10^{12}\mbox{cm}^{-2}$)}
 }
\startdata
$1-1$ & 23694.5 & $-$32 & 7 & $-$35 & 2 & 33 & 5 & $-$0.9 & 0.1 & 49 & 3 \\
$2-2$ & 23722.6 & $-$26 & 7 & $-$30 & 2 & 16 & 4 & $-$0.48 & 0.09 & 19 & 1 \\
$3-3$ & 23870.1 & $-$30 & 7 & $-$35 & 1 & 19 & 2 & $-$0.66 & 0.08 & 23 & 1 \\
$4-4$ & 24139.4 & $-$12 & 7 & $-$36 & 6 & 30 & 10 & $-$0.4 & 0.2 & 14 & 2 \\
$5-5$ & 24533.0 & \nodata & \nodata & \nodata & \nodata & \nodata & \nodata & \nodata & \nodata & \nodata & \nodata \\
$6-6$ & 25056.0 & $-$14 & 7 & $-$30 & 10 & 80 & 30 & $-$0.6 & 0.2 & 20 & 2 \\
\cline{1-12}\\
\colhead{Transition} & \colhead{Frequency} & \multicolumn{10}{c}{Narrow Absorption Features$^\dagger$}\\
\cline{3-12}
 & & \colhead{$T_A^*$} & \colhead{$\delta_{T_A^*}$} & \colhead{$V_{\tiny{\mbox{LSR}}}$} & \colhead{$\delta_{V_{\tiny{\mbox{LSR}}}}$} & \colhead{$\Delta_{\tiny{\mbox{FWHM}}}$} & \colhead{$\delta_{\Delta_{\tiny{\mbox{FWHM}}}}$} & \colhead{Area} & \colhead{$\delta_{\tiny{\mbox{Area}}}$} & \colhead{$N_{J,K}$} & \colhead{$\delta_{N_{J,K}}$}\\
\colhead{$J-K$} & & \colhead{(mK)} & \colhead{($\pm$mK)} & \colhead{(km/s)} & \colhead{($\pm$km/s)} & \colhead{(km/s)} & \colhead{($\pm$km/s)} & \colhead{(K km/s)} & \colhead{($\pm$K km/s)} & \colhead{($10^{12}\mbox{cm}^{-2}$)} & \colhead{($\pm 10^{12}\mbox{cm}^{-2}$)} \\
\cline{1-12} \\
$1-1$ & 23694.5 & \nodata & \nodata & \nodata & \nodata & \nodata & \nodata & \nodata & \nodata & \nodata & \nodata \\
$2-2$ & 23722.6 & $-$31 & 7 & $-$6: & \nodata & 2: & \nodata & $-$0.2: & \nodata & 6 & \nodata \\
$3-3$ & 23870.1 & $-$45 & 7 & $-$6.4 & 0.4 & 6.1 & 0.8 & $-$0.37 & 0.04 & 13 & 1 \\
$4-4$ & 24139.4 & $-$41 & 7 & $-$5 & 1 & 4 & 1 & $-$0.23 & 0.06 & 7 & 1 \\
$5-5$ & 24533.0 & $-$44 & 7 & $-$4.6 & 0.4 & 4.7 & 0.7 & $-$0.27 & 0.04 & 8.4 & 0.5 \\
$6-6$ & 25056.0 & $-$43 & 7 & $-$4.5 & 0.5 & 5 & 1 & $-$0.25 & 0.04 & 8 & 1
\enddata

\tablenotetext{\dagger}{\,Marginal detections are annotated with a colon (:).}

%\tablenotetext{a}{Frequencies of unresolved hyperfine structures centered at the strongest pertinent emission line listed in the NIST Recommended Rest Frequencies for Observed Interstellar Molecular Microwave Transitions \citep{lov04}.}
\end{deluxetable}

\begin{deluxetable}{lcccc}
%\rotate
\tablecaption{Excitation temperature and column density of ammonia molecules detected in the 20--25\,GHz NRO spectrum of CRL\,618.
\label{tbl:618_rot}}
\tabletypesize{\scriptsize}
\tablewidth{0pt}
\tablehead{
\colhead{Component} & \colhead{$T_{ex}$} & \colhead{$\delta_{T_{ex}}$} &\colhead{$N$} & \colhead{$\delta_{N}$} \\
 & \colhead{(K)} & \colhead{($\pm$K)} & \colhead{($10^{14}$ cm$^{-2}$)} & \colhead{($\pm10^{14}$ cm$^{-2}$)} }
\startdata
Broad absorption (low $T_{ex}$ region) & 28 & 10 & 3.6 & 0.5 \\
Broad absorption (high $T_{ex}$ region) & 400 & 200 & 10 & 2 \\
Narrow absorption & 290 & 70 & 4.4 & 0.8
\enddata
\end{deluxetable}

\clearpage

\begin{figure}
	\centering
	\begin{subfigure}
		\centering
		\includegraphics[width=\textwidth]{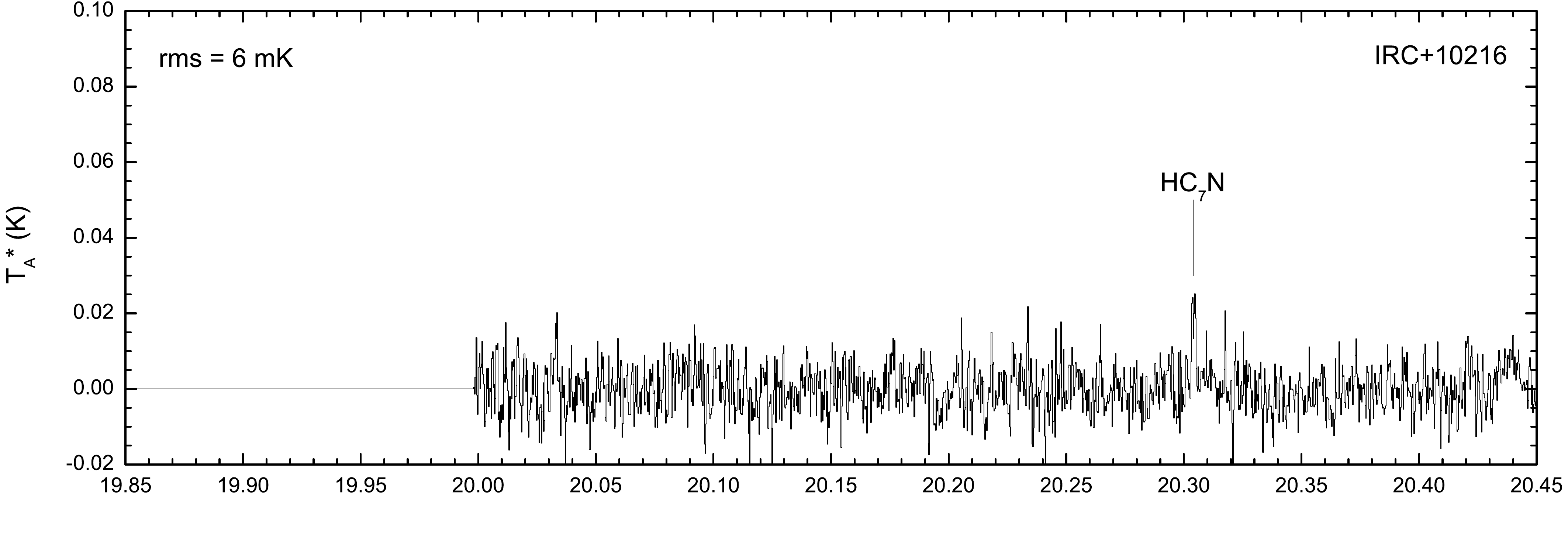}
	\end{subfigure}
	\begin{subfigure}
		\centering
		\includegraphics[width=\textwidth]{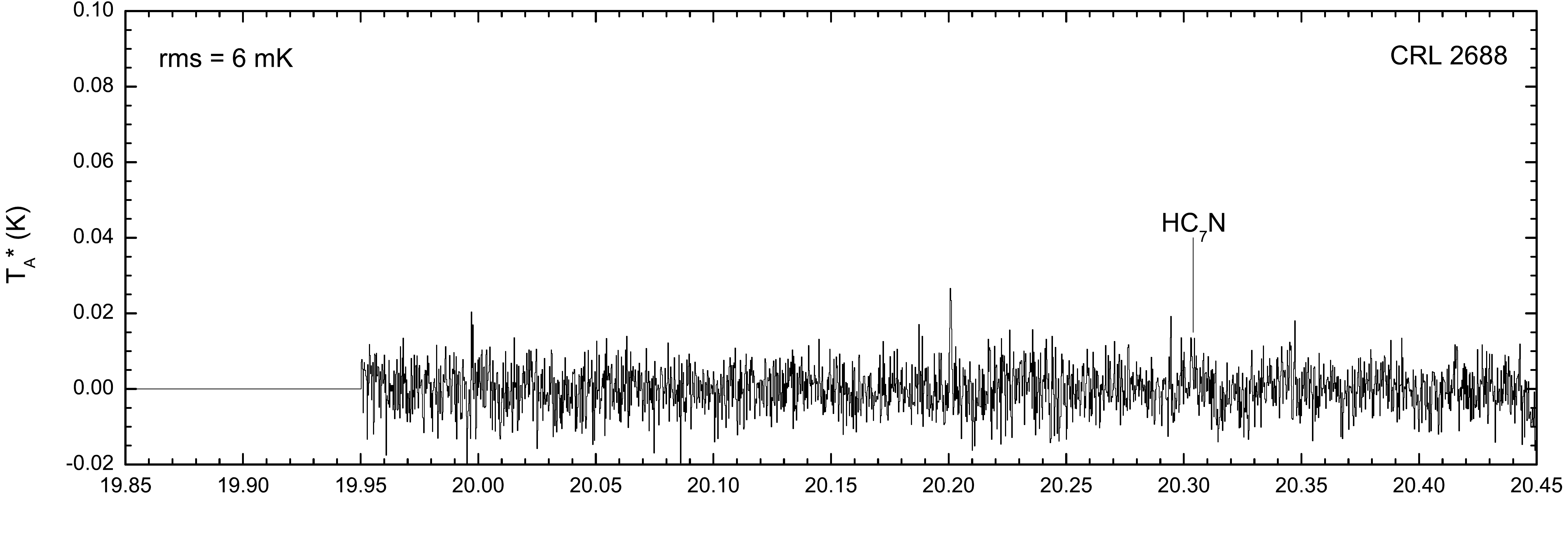}
	\end{subfigure}
	\begin{subfigure}
		\centering
		\includegraphics[width=\textwidth]{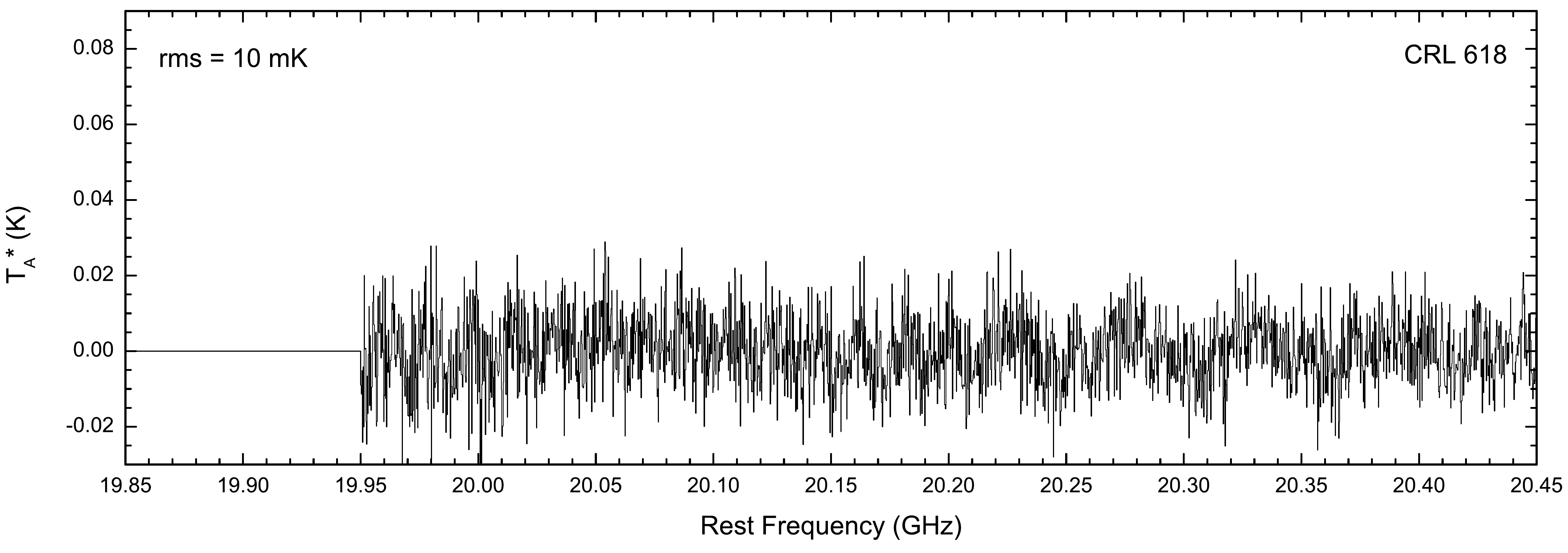}
	\end{subfigure}
	\caption{The reduced 20--25\,GHz NRO spectra of IRC+10216, CRL\,2688, CRL\,618, and NGC\,7027. The rms level of each region is shown. Positively identified features have also been labelled for ease of reference. 
}\label{fig:zoom}
\end{figure}

\begin{figure}
	\setcounter{figure}{1}
%	\ContinuedFloat
	\begin{subfigure}
		\centering
		\includegraphics[width=\textwidth]{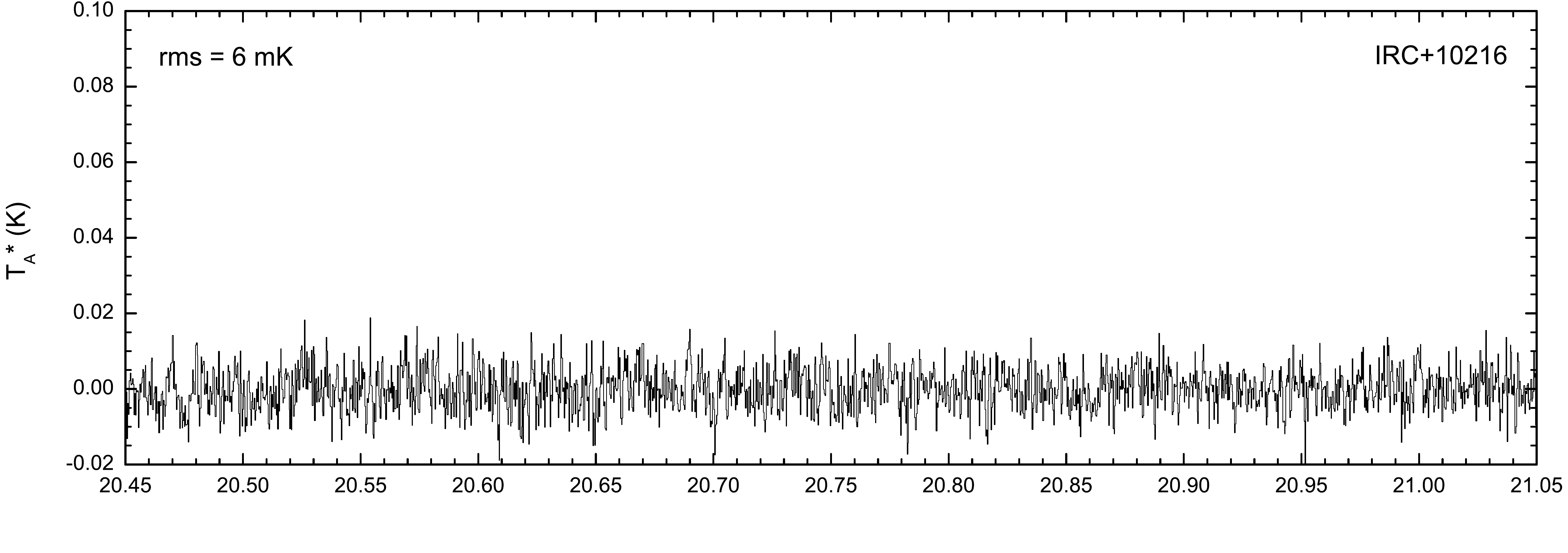}
	\end{subfigure}
	\begin{subfigure}
		\centering
		\includegraphics[width=\textwidth]{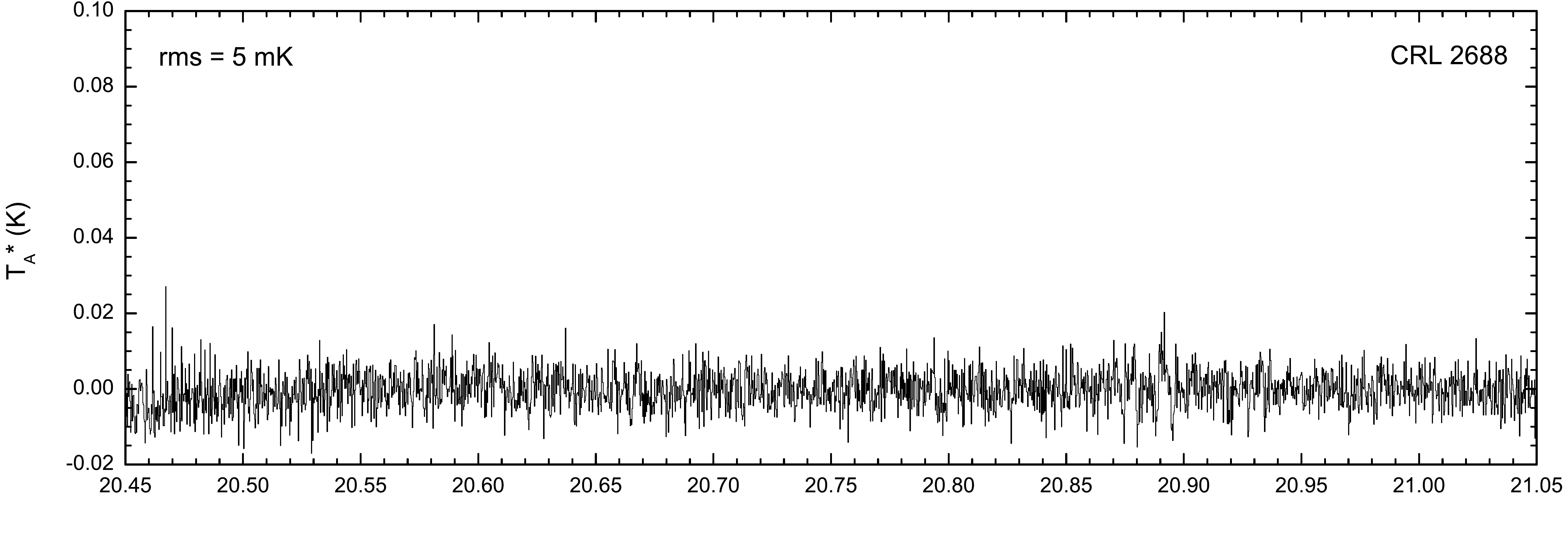}
	\end{subfigure}
	\begin{subfigure}
		\centering
		\includegraphics[width=\textwidth]{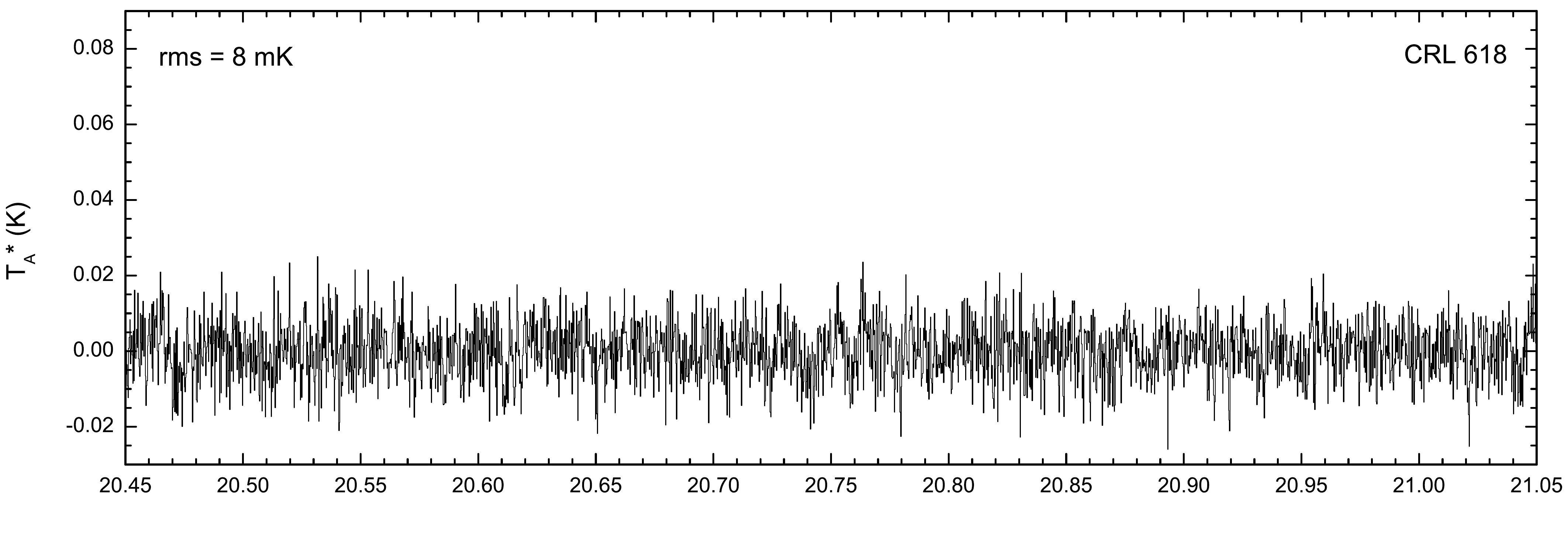}
	\end{subfigure}
	\begin{subfigure}
		\centering
		\includegraphics[width=\textwidth]{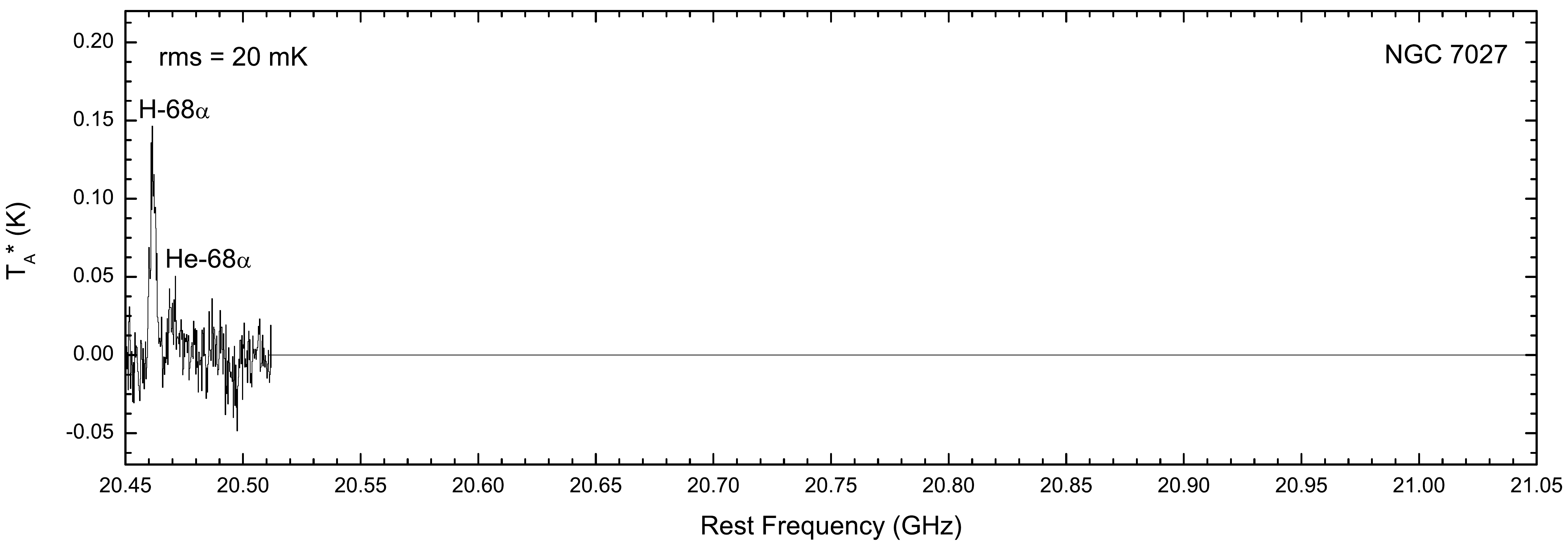}
	\end{subfigure}
	\caption{(continued)}
\end{figure}

\begin{figure}
	\setcounter{figure}{1}
%	\ContinuedFloat
	\begin{subfigure}
		\centering
		\includegraphics[width=\textwidth]{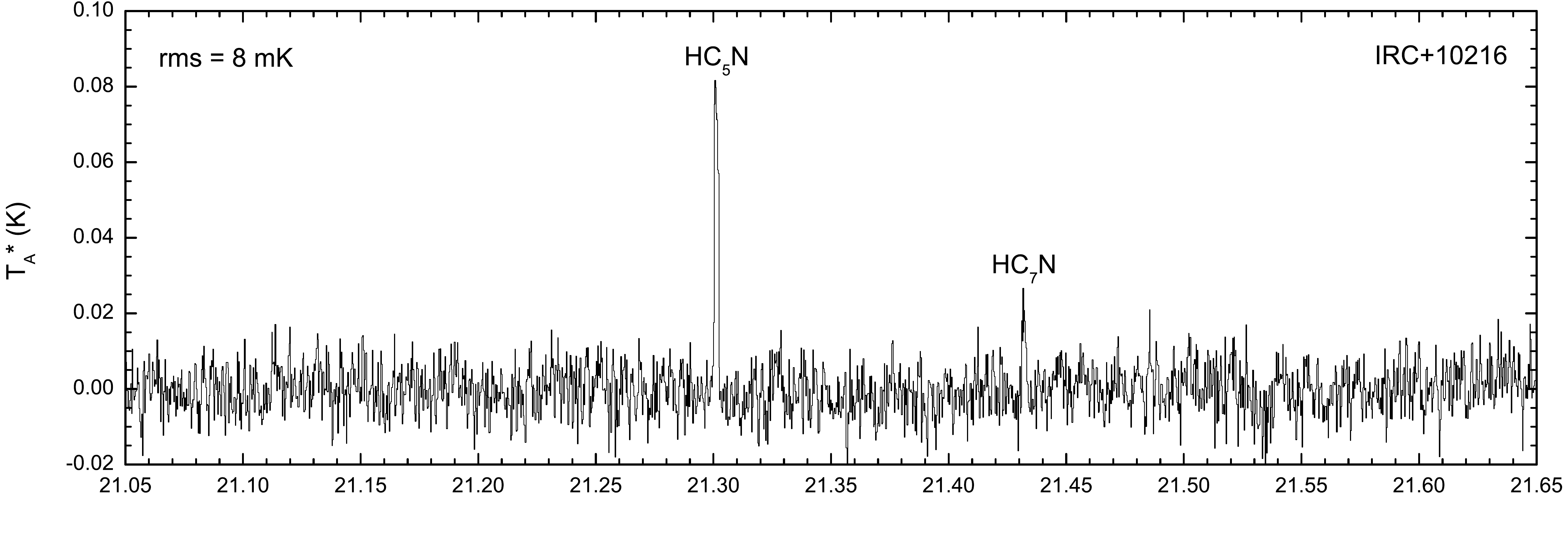}
	\end{subfigure}
	\begin{subfigure}
		\centering
		\includegraphics[width=\textwidth]{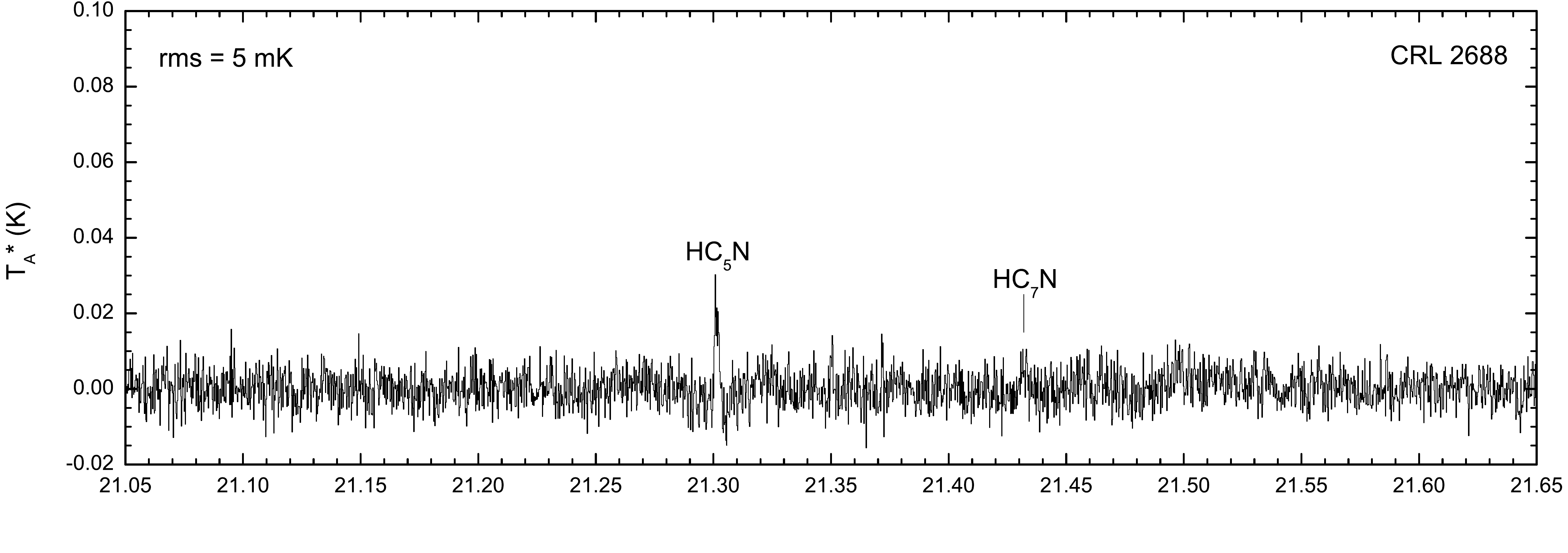}
	\end{subfigure}
	\begin{subfigure}
		\centering
		\includegraphics[width=\textwidth]{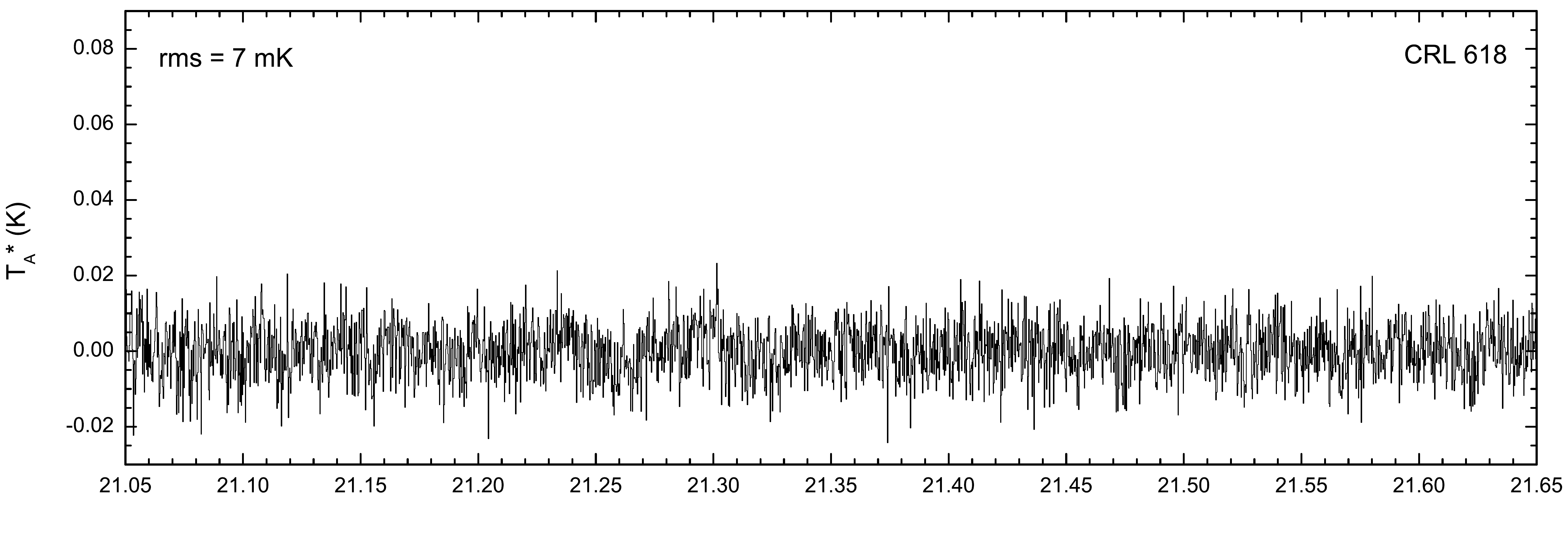}
	\end{subfigure}
	\begin{subfigure}
		\centering
		\includegraphics[width=\textwidth]{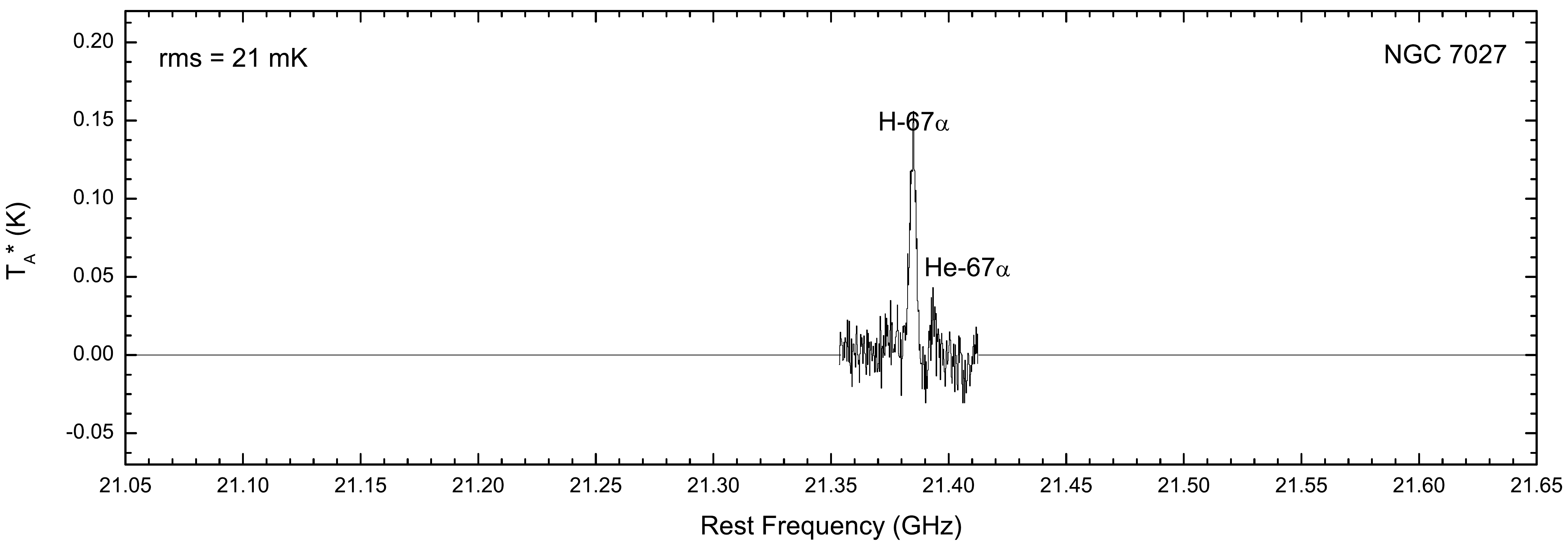}
	\end{subfigure}
	\caption{(continued)}
\end{figure}

\begin{figure}
	\setcounter{figure}{1}
%	\ContinuedFloat
	\begin{subfigure}
		\centering
		\includegraphics[width=\textwidth]{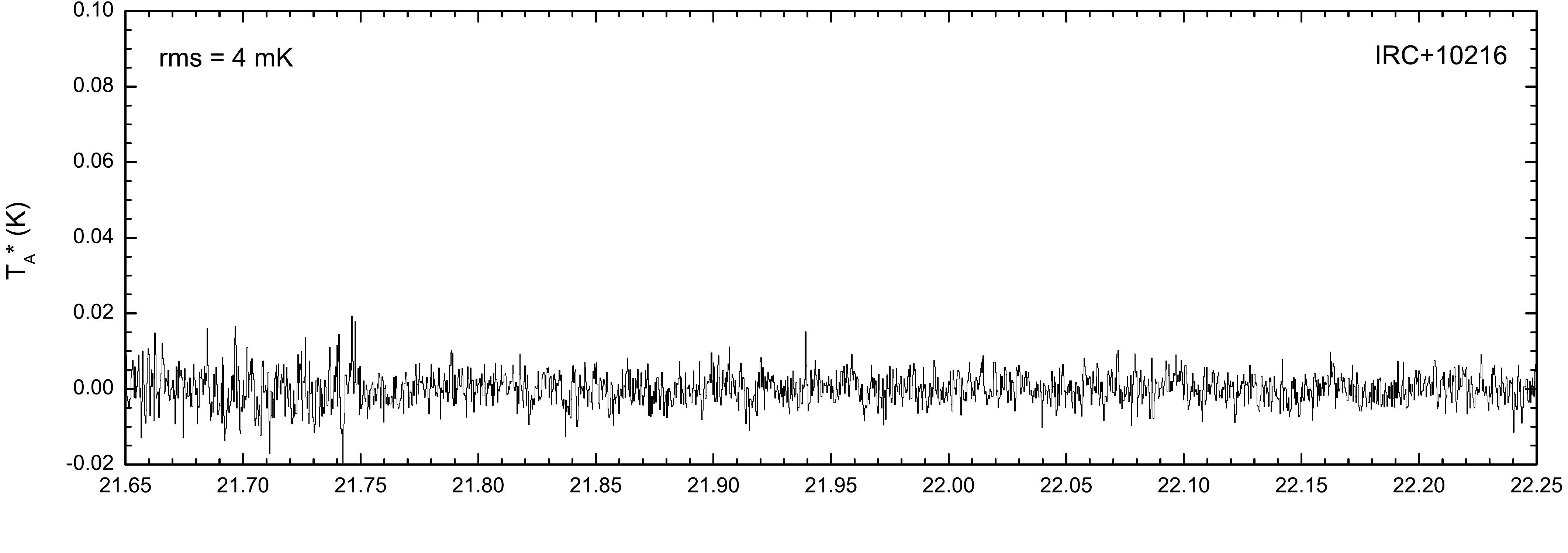}
	\end{subfigure}
	\begin{subfigure}
		\centering
		\includegraphics[width=\textwidth]{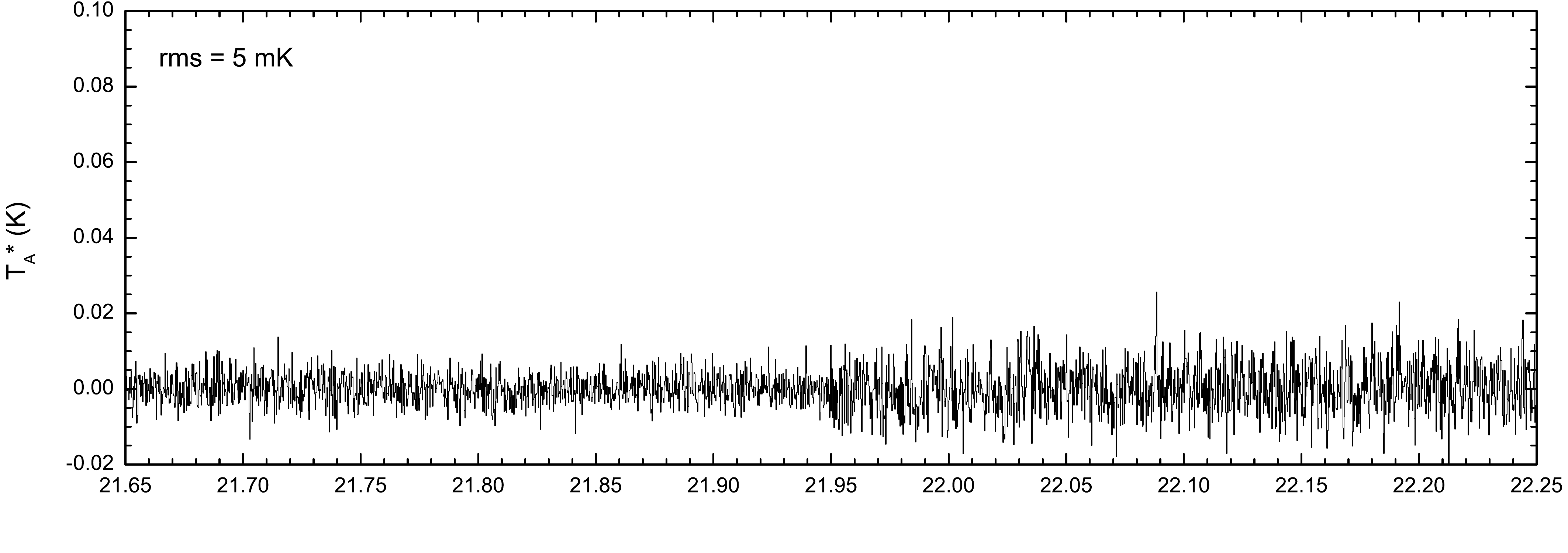}
	\end{subfigure}
	\begin{subfigure}
		\centering
		\includegraphics[width=\textwidth]{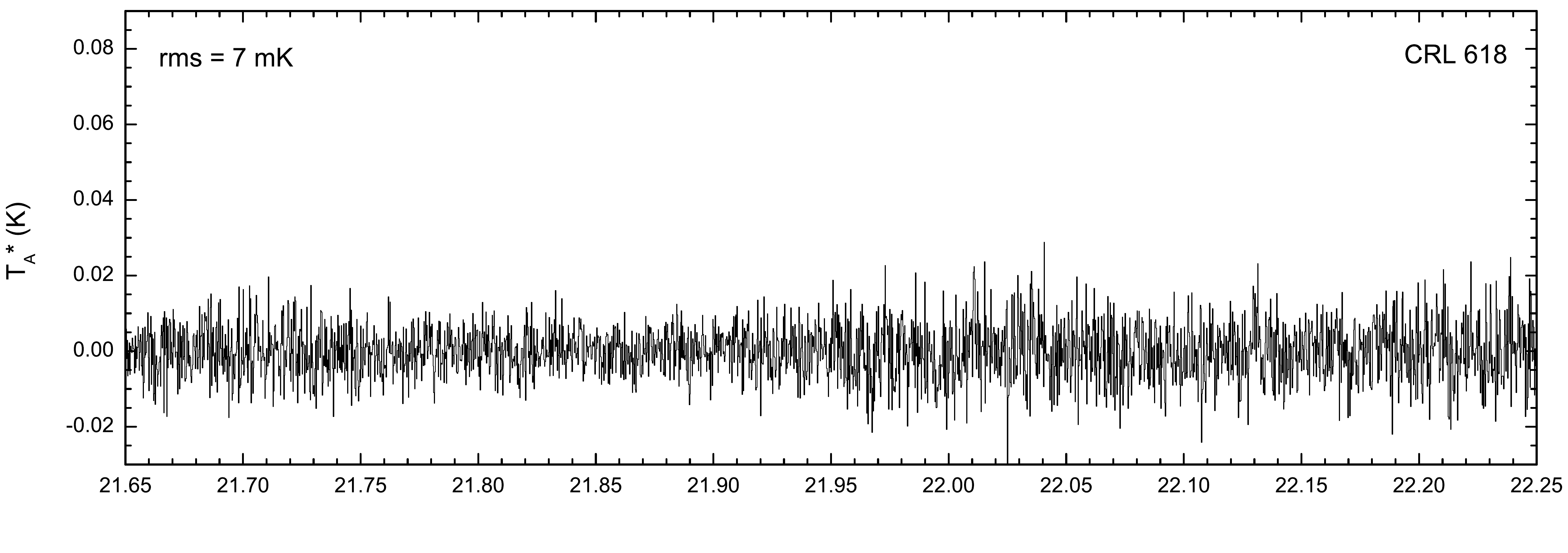}
	\end{subfigure}
	\begin{subfigure}
		\centering
		\includegraphics[width=\textwidth]{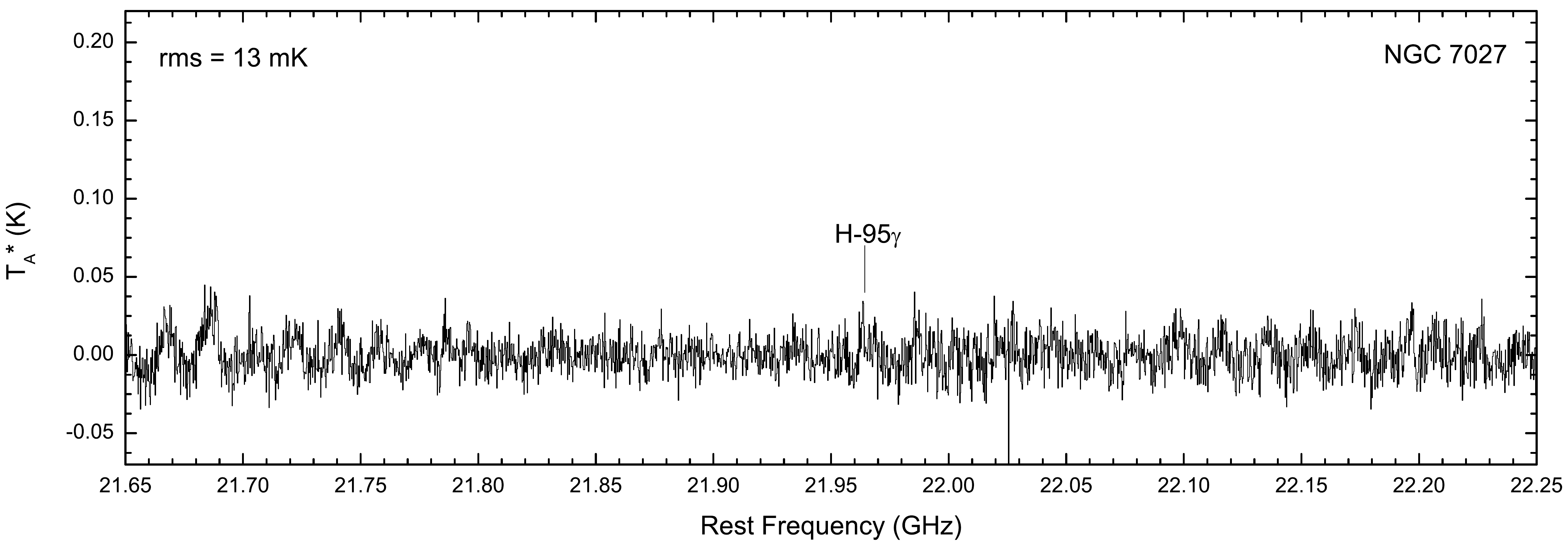}
	\end{subfigure}
	\caption{(continued)}
\end{figure}

\begin{figure}
	\setcounter{figure}{1}
%	\ContinuedFloat
	\begin{subfigure}
		\centering
		\includegraphics[width=\textwidth]{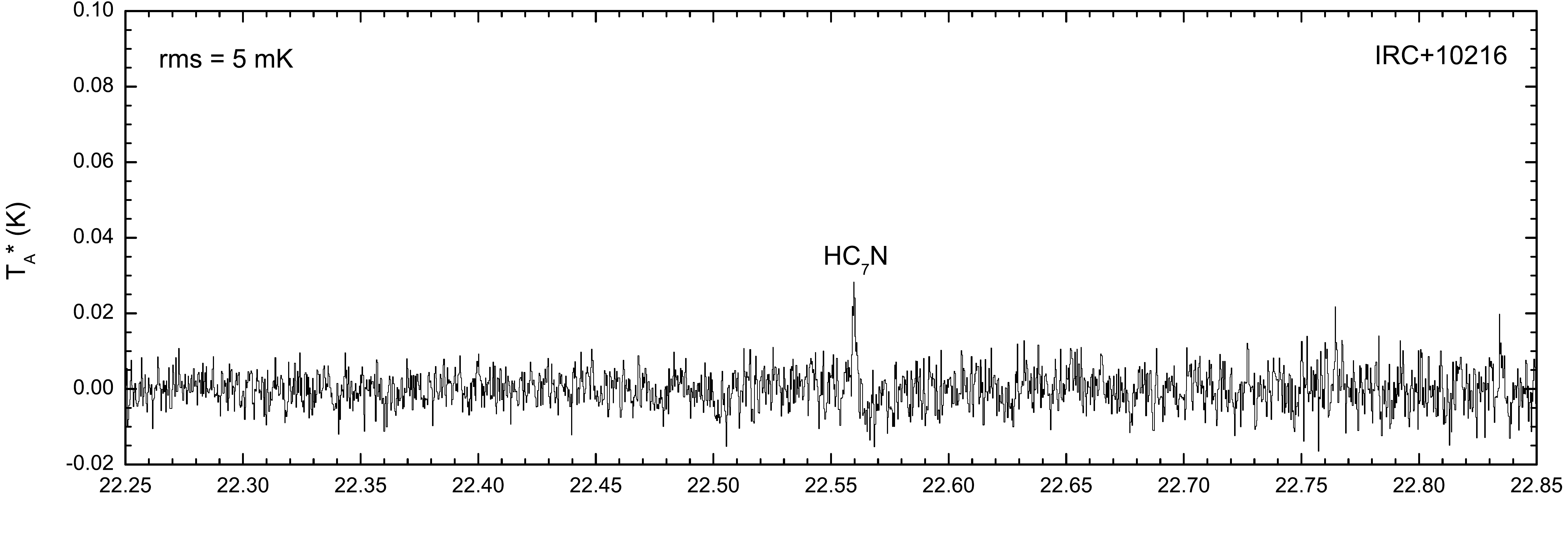}
	\end{subfigure}
	\begin{subfigure}
		\centering
		\includegraphics[width=\textwidth]{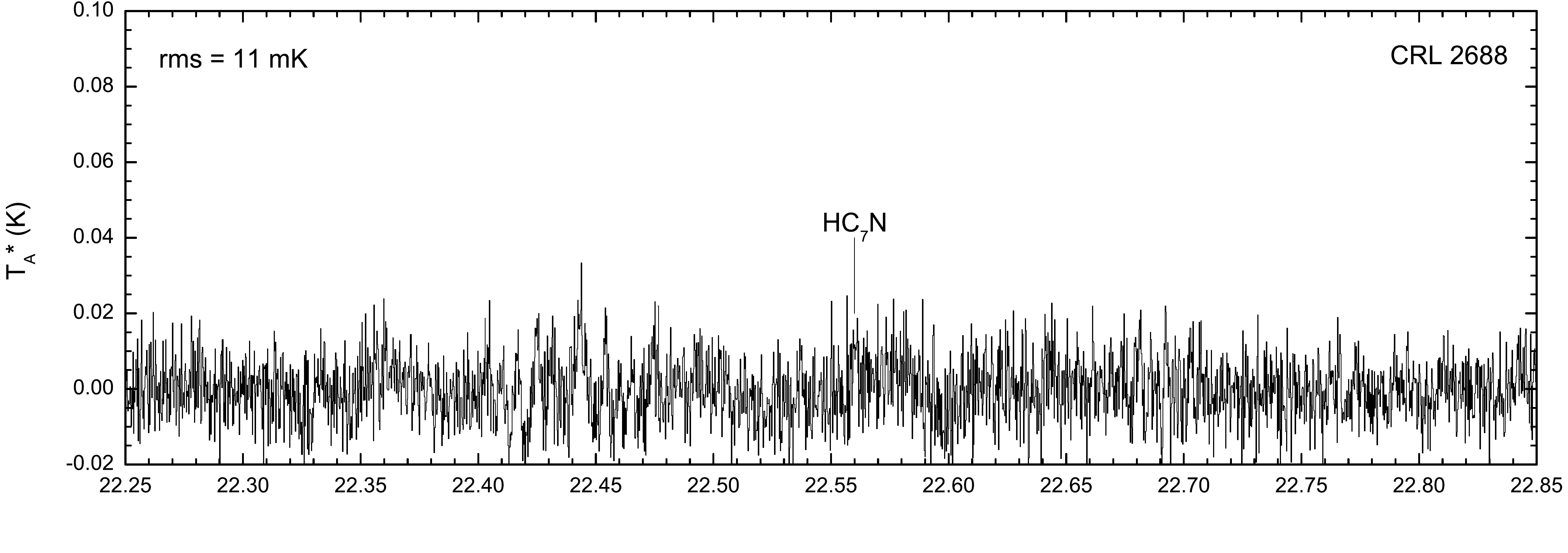}
	\end{subfigure}
	\begin{subfigure}
		\centering
		\includegraphics[width=\textwidth]{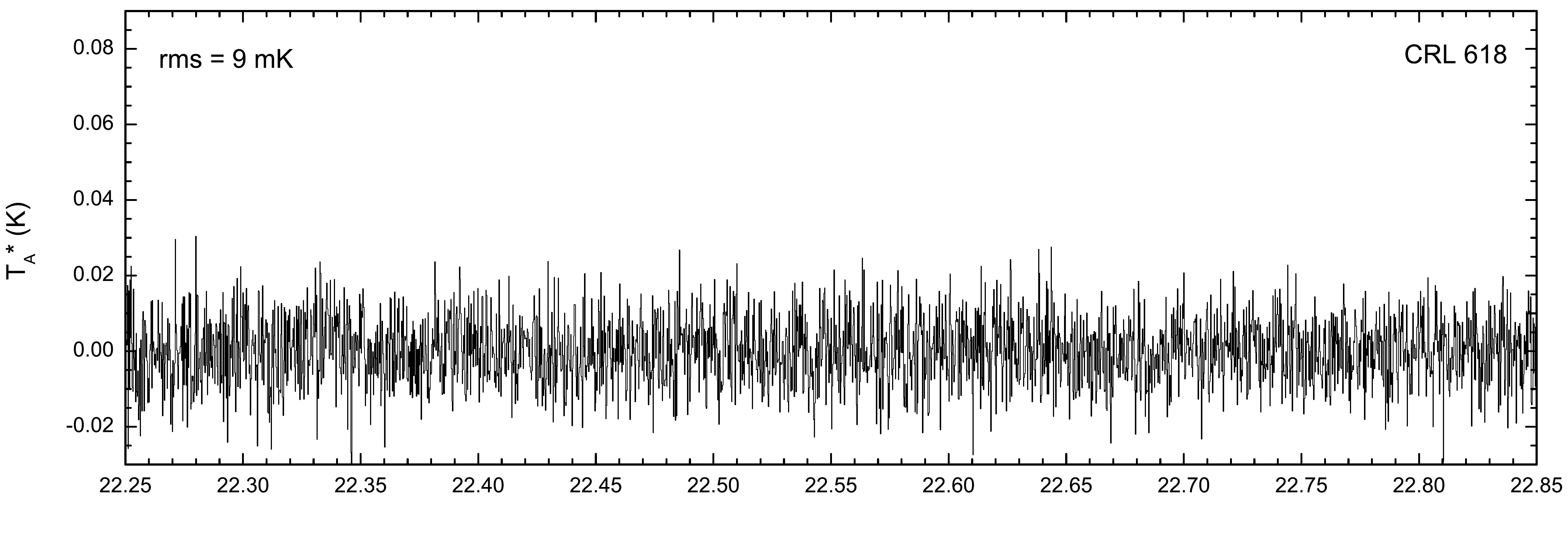}
	\end{subfigure}
	\begin{subfigure}
		\centering
		\includegraphics[width=\textwidth]{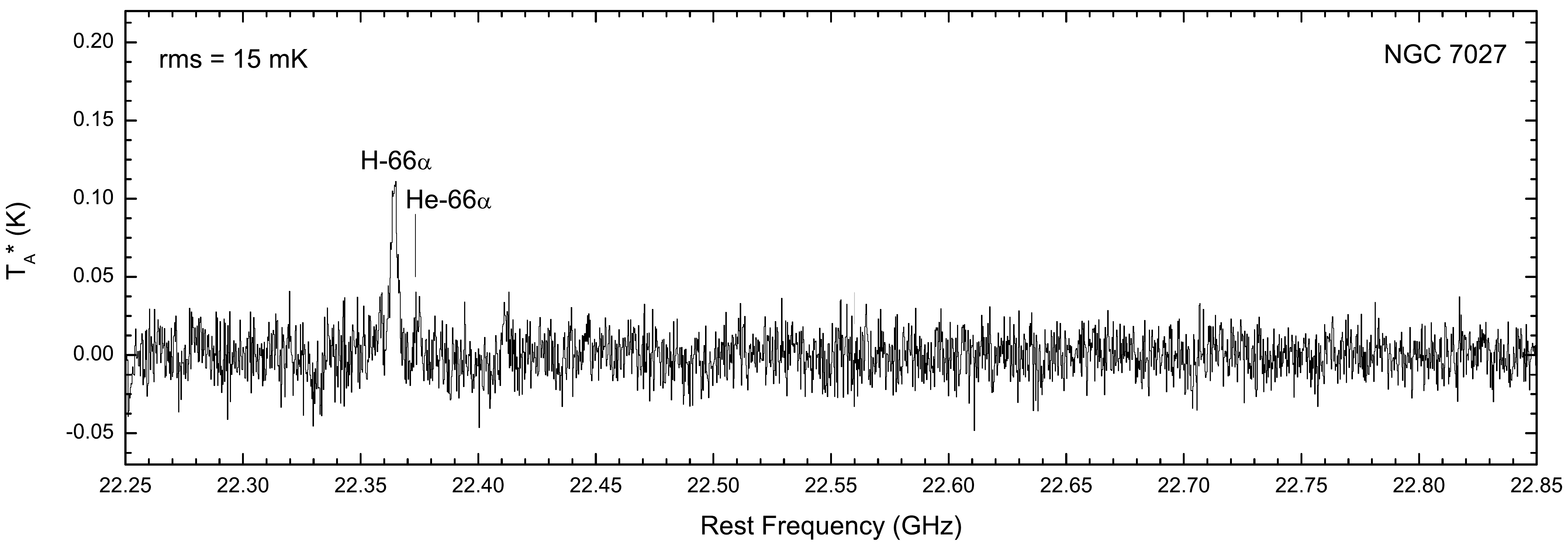}
	\end{subfigure}
	\caption{(continued)}
\end{figure}

\begin{figure}
	\setcounter{figure}{1}
%	\ContinuedFloat
	\begin{subfigure}
		\centering
		\includegraphics[width=\textwidth]{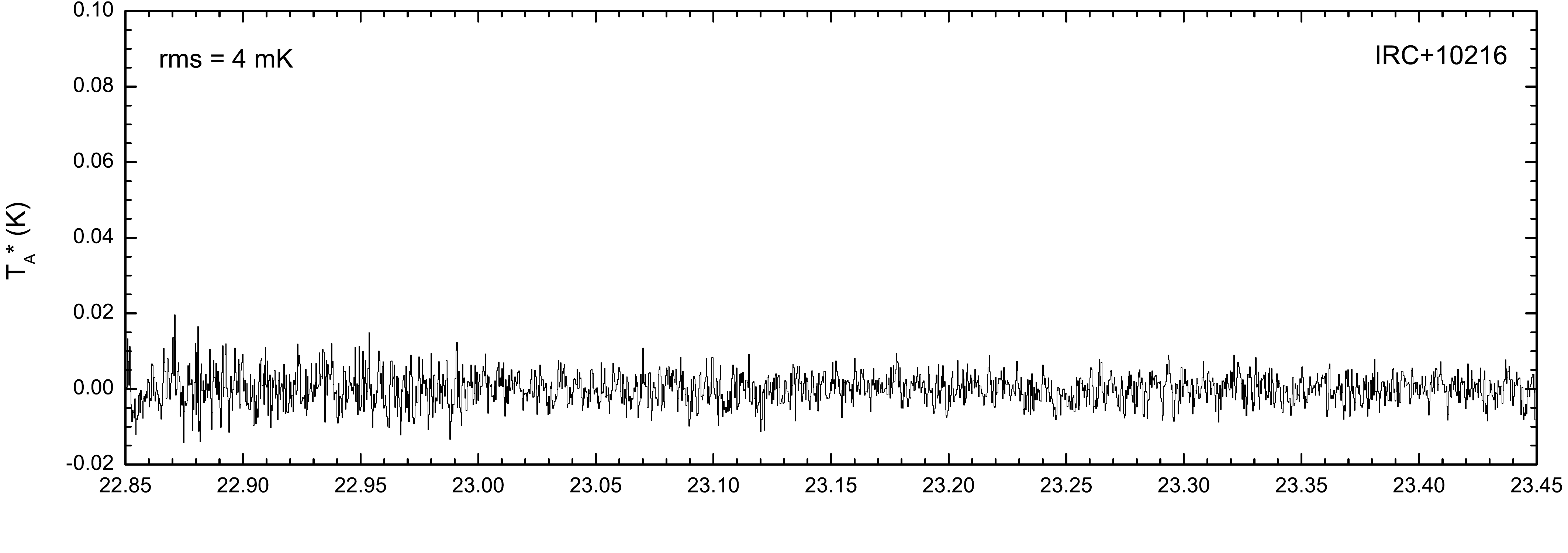}
	\end{subfigure}
	\begin{subfigure}
		\centering
		\includegraphics[width=\textwidth]{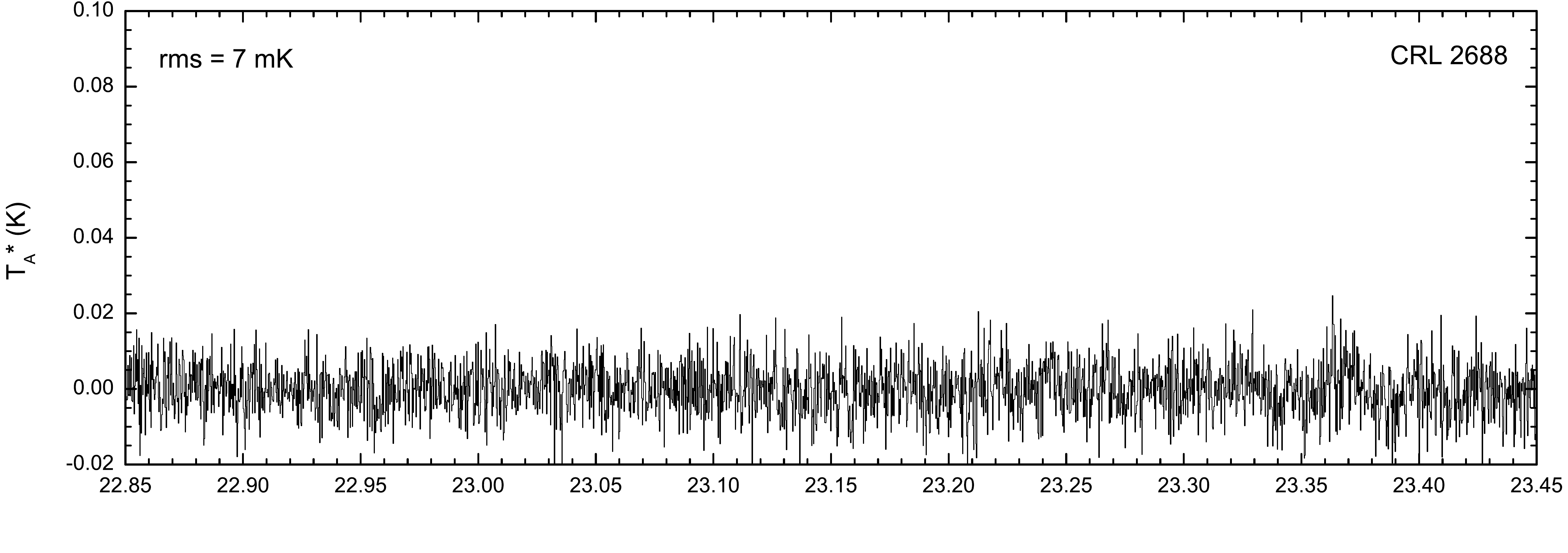}
	\end{subfigure}
	\begin{subfigure}
		\centering
		\includegraphics[width=\textwidth]{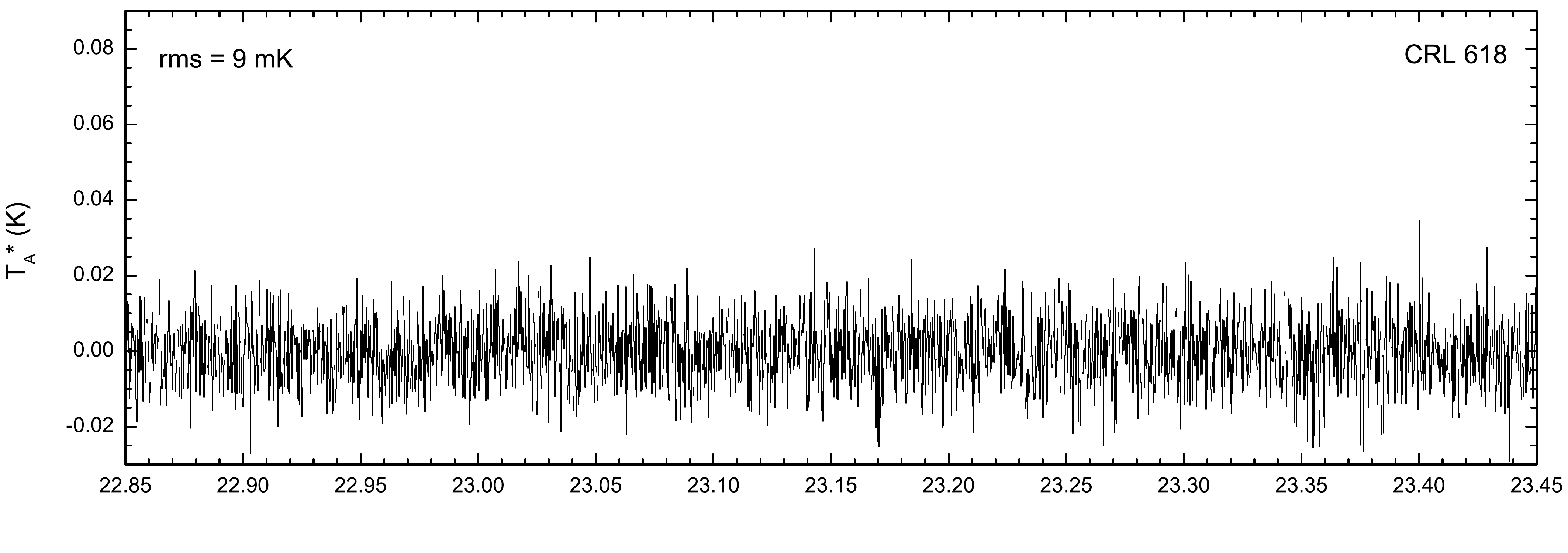}
	\end{subfigure}
	\begin{subfigure}
		\centering
		\includegraphics[width=\textwidth]{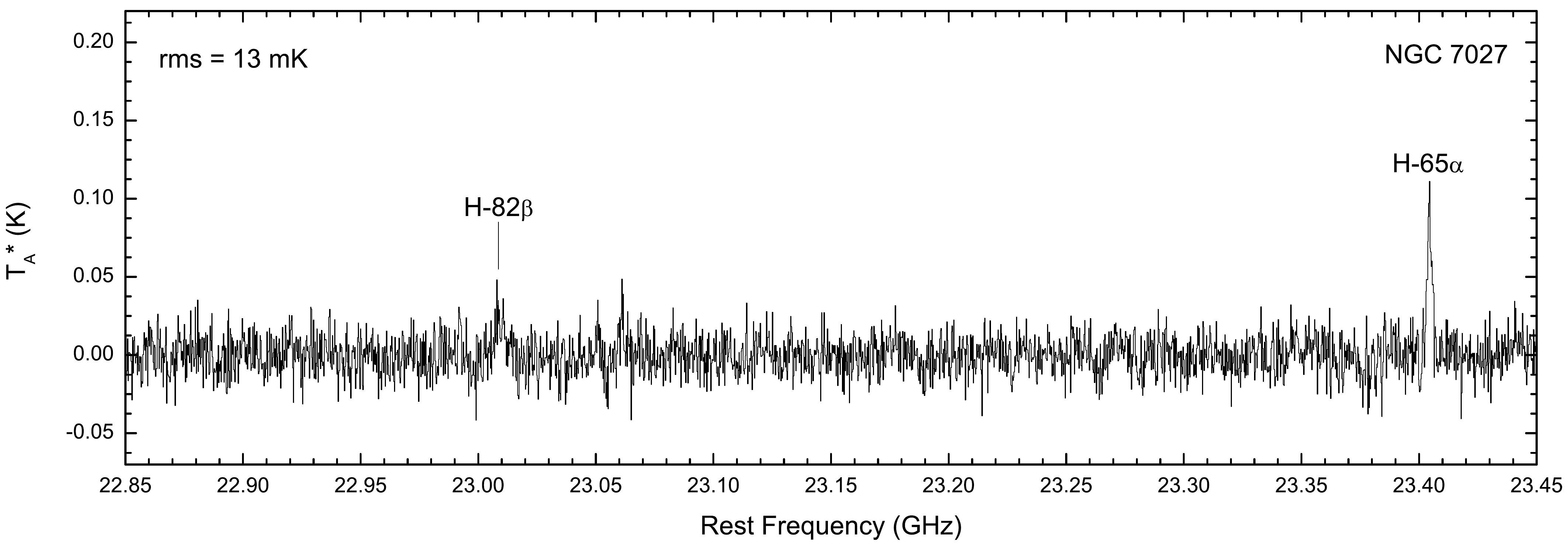}
	\end{subfigure}
	\caption{(continued)}
\end{figure}

\begin{figure}
	\setcounter{figure}{1}
%	\ContinuedFloat
	\begin{subfigure}
		\centering
		\includegraphics[width=\textwidth]{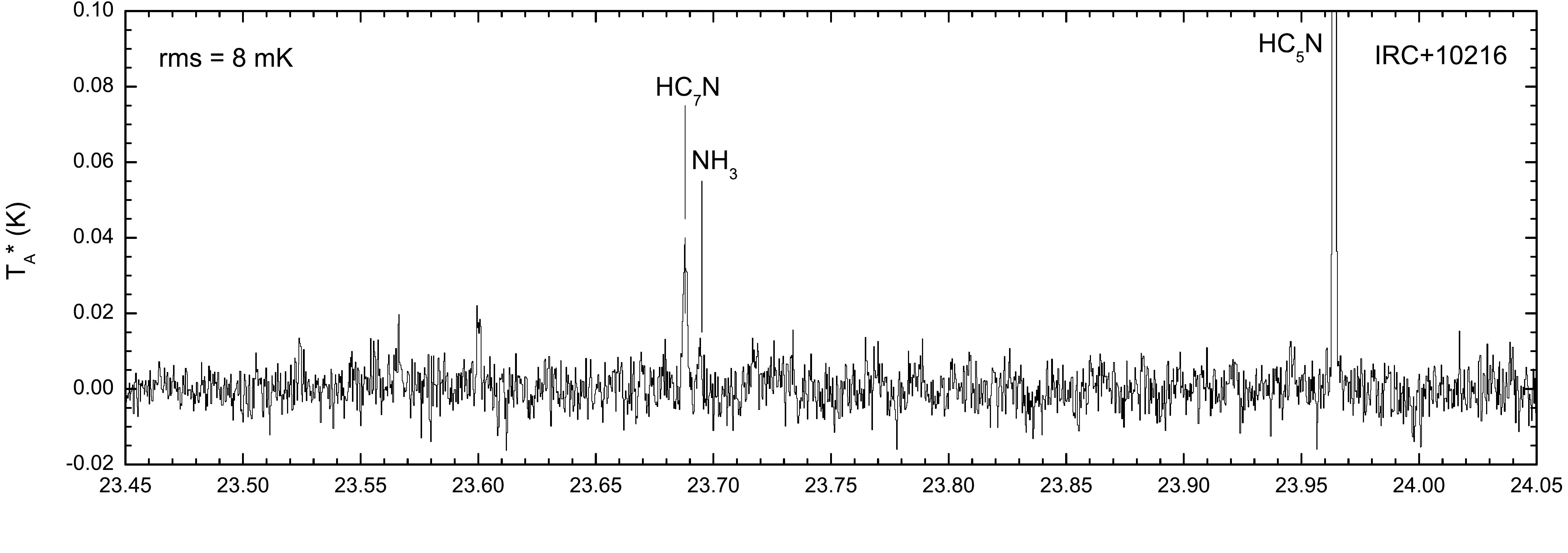}
	\end{subfigure}
	\begin{subfigure}
		\centering
		\includegraphics[width=\textwidth]{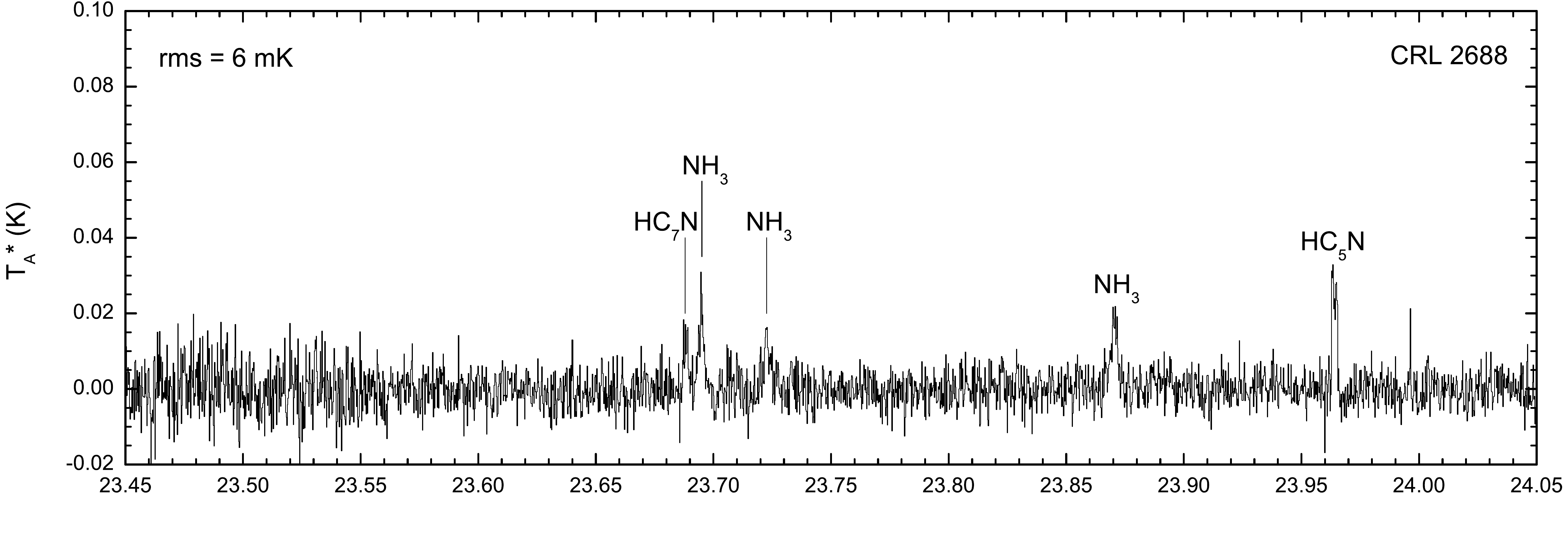}
	\end{subfigure}
	\begin{subfigure}
		\centering
		\includegraphics[width=\textwidth]{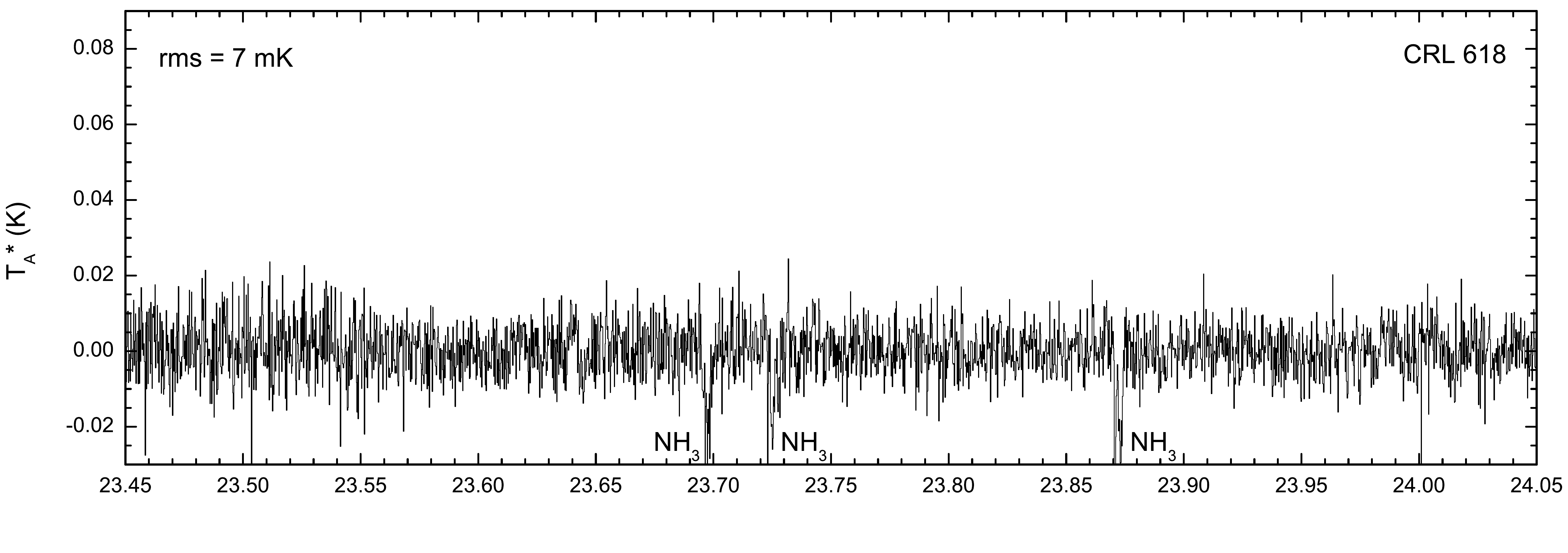}
	\end{subfigure}
	\begin{subfigure}
		\centering
		\includegraphics[width=\textwidth]{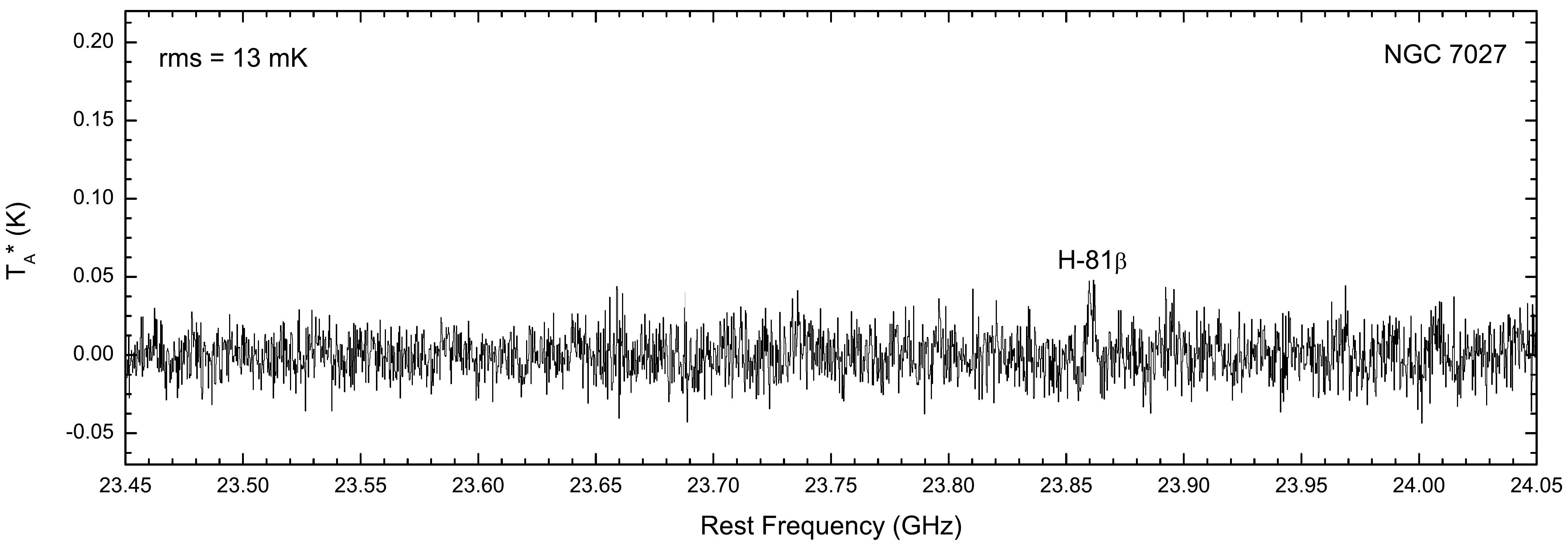}
	\end{subfigure}
	\caption{(continued)}
\end{figure}

\begin{figure}
	\setcounter{figure}{1}
%	\ContinuedFloat
	\begin{subfigure}
		\centering
		\includegraphics[width=\textwidth]{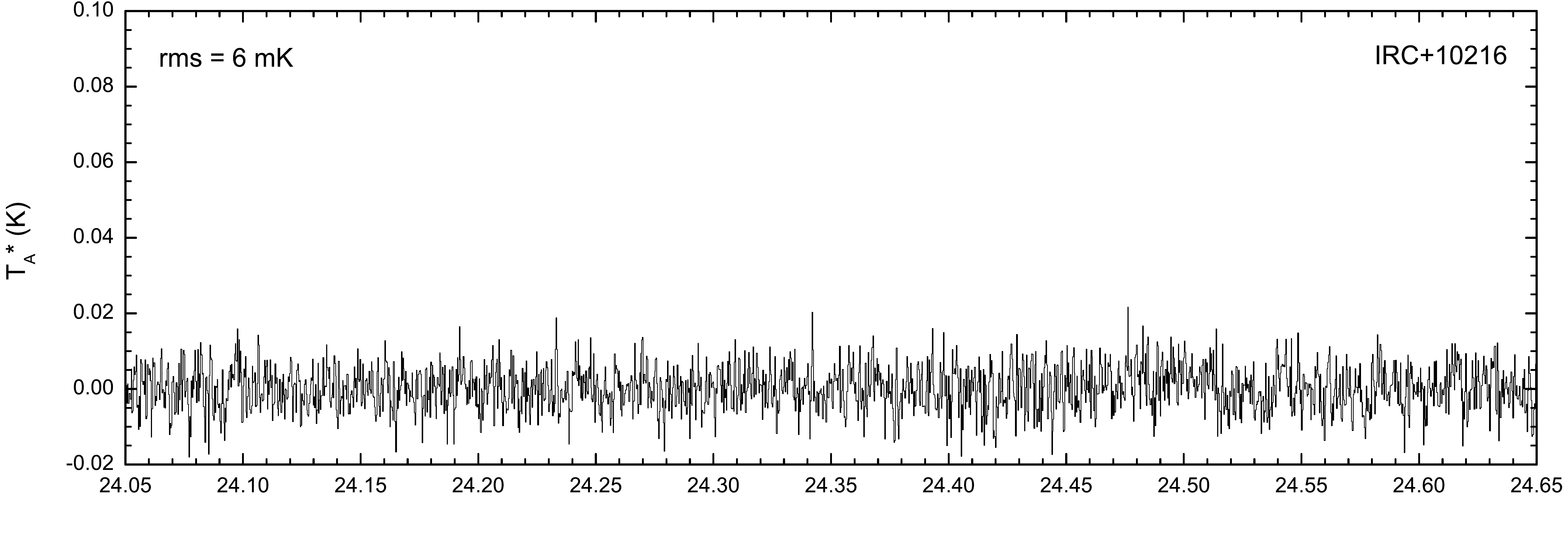}
	\end{subfigure}
	\begin{subfigure}
		\centering
		\includegraphics[width=\textwidth]{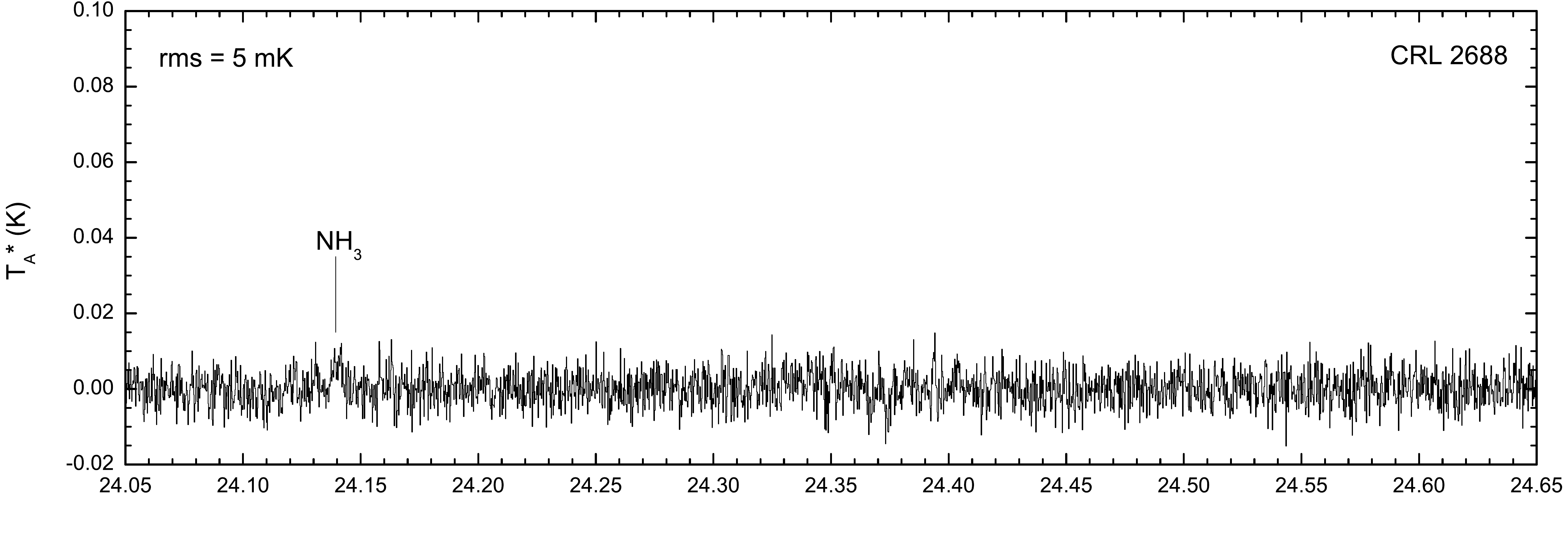}
	\end{subfigure}
	\begin{subfigure}
		\centering
		\includegraphics[width=\textwidth]{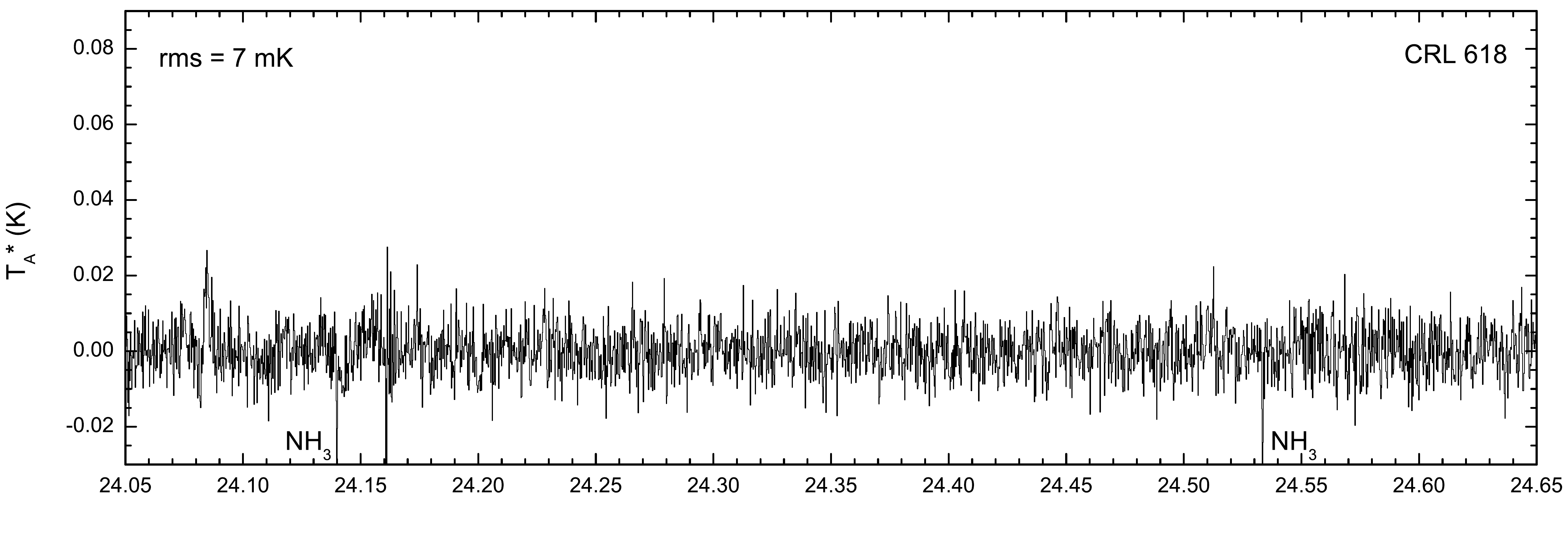}
	\end{subfigure}
	\begin{subfigure}
		\centering
		\includegraphics[width=\textwidth]{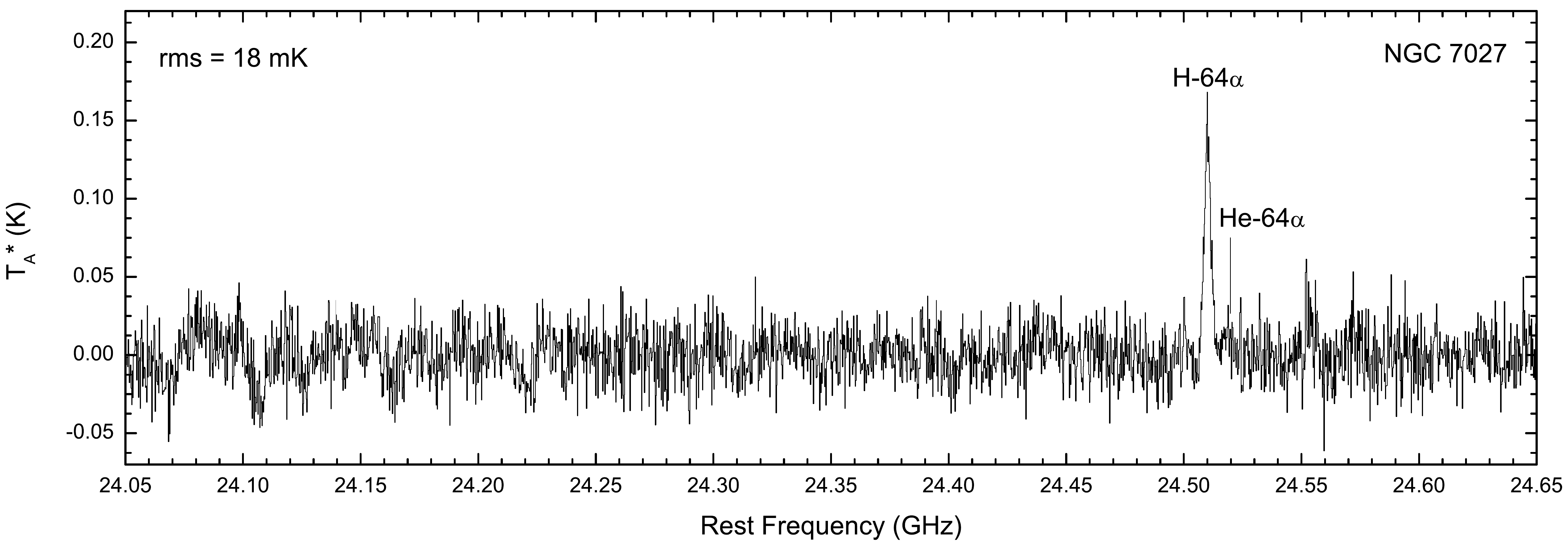}
	\end{subfigure}
	\caption{(continued)}
\end{figure}

\begin{figure}
	\setcounter{figure}{1}
%	\ContinuedFloat
	\begin{subfigure}
		\centering
		\includegraphics[width=\textwidth]{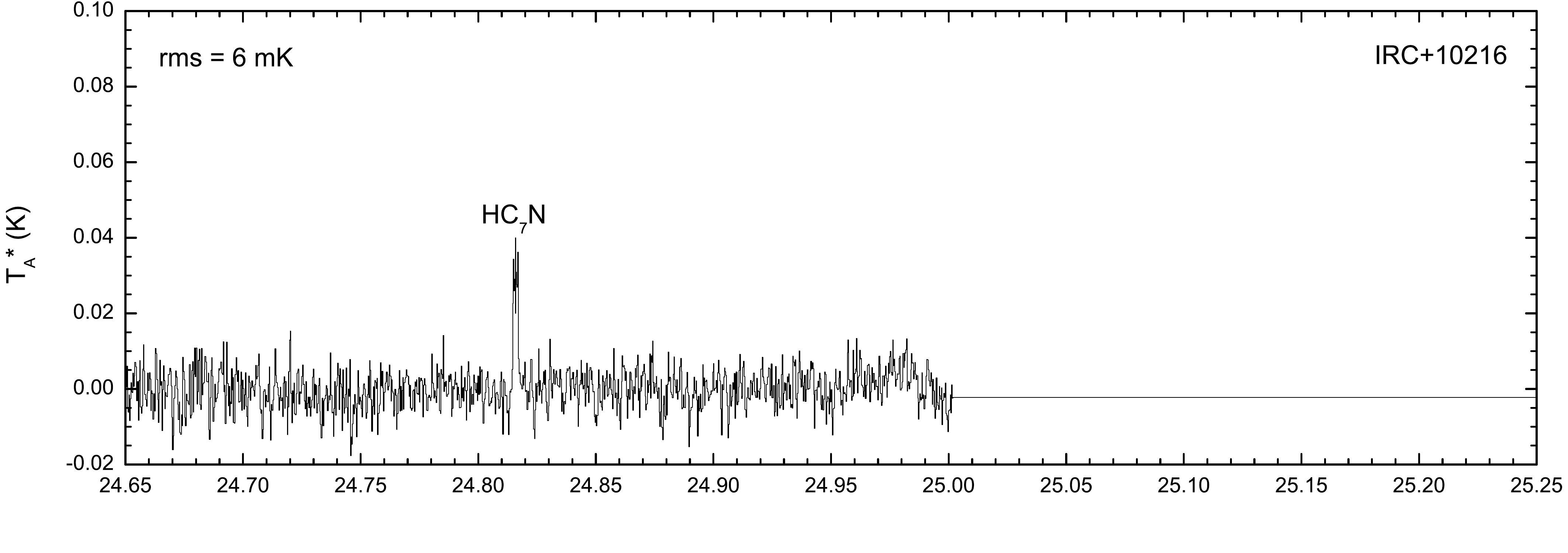}
	\end{subfigure}
	\begin{subfigure}
		\centering
		\includegraphics[width=\textwidth]{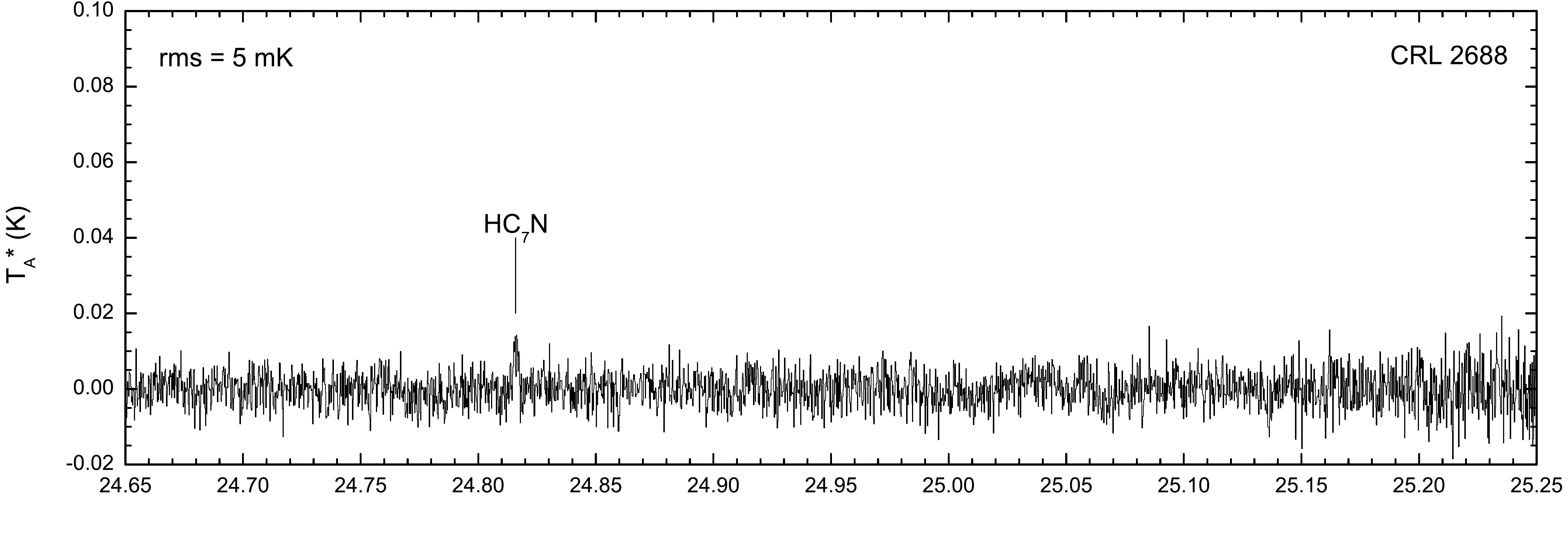}
	\end{subfigure}
	\begin{subfigure}
		\centering
		\includegraphics[width=\textwidth]{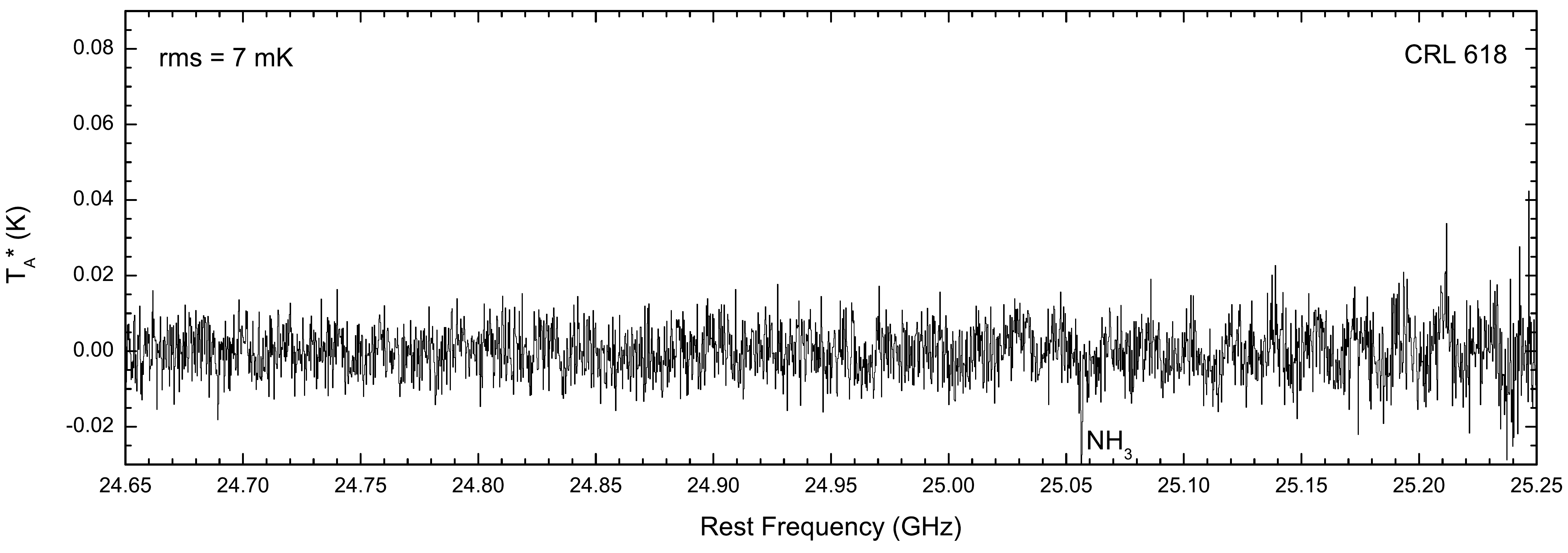}
	\end{subfigure}
	\caption{(continued)}
\end{figure}

%\begin{figure}
%	\centering
%	\includegraphics[width=\textwidth]{figures/7027_ratio/7027_ratio.eps}
%	\caption{Flux ratios of H$n\alpha$ recombinatures features relative to the H39$\alpha$ line in NGC\,7027. Theortical models with $T_e=1.25\times10^4\,\mbox{K}$ and $N_e=10^4\,\mbox{cm}^{-3}$ (\textit{red line}), $T_e=1.00\times10^4\,\mbox{K}$ and $N_e=10^4\,\mbox{cm}^{-3}$ (\textit{green line}), and $T_e=1.25\times10^4\,\mbox{K}$ and $N_e=10^3\,\mbox{cm}^{-3}$ (\textit{blue line}) from \citet{sto95} have also been plotted.}\label{fig:7027_ratio}
%\end{figure}

\end{document}